\documentclass[final,3p,times]{elsarticle}

\usepackage{amssymb,amsmath,amsfonts}
\usepackage{tabularx} 
\usepackage{xcolor}
\usepackage{algorithm}
\usepackage[boxruled,algo2e]{algorithm2e}
\usepackage[hidelinks]{hyperref}

\usepackage{hyperref}
\usepackage{soul}
\usepackage{subcaption}
\usepackage{booktabs}

\usepackage{natbib}
\usepackage{tikz}
\usetikzlibrary{patterns}
\usepackage[precision=2, unit=mm]{lengthconvert}

\begin{document}

\begin{frontmatter}
\title{An AMReX-based Compressible Reacting Flow Solver for High-speed Reacting Flows relevant to Hypersonic Propulsion}

\author[inst1]{Shivank Sharma}
\author[inst1]{Ral Bielawski}
\author[inst1]{Oliver Gibson}
\author[inst1]{Shuzhi Zhang}
\author[inst1]{Vansh Sharma}
\author[inst1]{Andreas H. Rauch}
\author[inst1]{Jagmohan Singh}
\author[inst1]{Sebastian Abisleiman}
\author[inst1]{Michael Ullman}
\author[inst1,inst2]{Shivam Barwey}
\author[inst1]{Venkat Raman}
\date{}

\cortext[cor1]{Corresponding author. E-mail address: \href{mailto:sshivank@umich.edu}{sshivank@umich.edu} (S.\,Sharma)}

\affiliation[inst1]{organization={Department of Aerospace Engineering, University of Michigan},
            addressline={1320 Beal Ave.}, 
            city={Ann Arbor},
            postcode={48109}, 
            state={MI},
            country={USA}}

\affiliation[inst2]{organization={Argonne Leadership Computing Facility, Argonne National Laboratory},
            addressline={9700 S Cass Ave.}, 
            city={Lemont},
            postcode={60439}, 
            state={IL},
            country={USA}}

\begin{abstract}
This work presents a comprehensive framework for the efficient implementation of finite-volume-based reacting flow solvers, specifically tailored for high speed propulsion applications. Using the exascale computing project (ECP) based AMReX framework, a compressible flow solver for handling high-speed reacting flows is developed. This work is complementary to the existing PeleC solver \cite{peleC}, emphasizing specific applications that include confined shock-containing flows, stationary and moving shocks and detonations. The framework begins with a detailed exposition of the numerical methods employed, emphasizing their application to complex geometries and their effectiveness in ensuring accurate and stable numerical simulations. Subsequently, an in-depth analysis evaluates the solver’s performance across canonical and practical geometries, with particular focus on computational cost and efficiency. The solver's scalability and robustness are demonstrated through practical test cases, including flow path simulations of scramjet engines and detailed analysis of various detonation phenomena. 
\end{abstract}
\end{frontmatter}

\tableofcontents 

\section{Introduction and motivation}

Computational modeling is at the cusp of a new era of computing, driven in large by single-instruction multiple-thread (SIMT) architectures epitomized by the general-purpose graphical processing units (GPUs). With the recent focus on artificial intelligence and machine learning, the low energy usage (per floating point operation) of GPUs made them attractive for compute-intensive applications \cite{huang2009energy}. In preparation for this exascale era, there has been a concerted effort in the development of algorithms and codes that will take advantage of the vast and heterogeneous machines that are becoming commonplace \cite{messina2017exascale}. The main focus has been the limits imposed by data transfer between GPU and CPU components of the machine \cite{AMReX_JOSS,grete2023parthenon,barwey2021neural}. Existing methods for large-scale computational physics have been developed mainly for CPU-based computing, where memory access limitations are far less pronounced. To reduce the energy budget of computing, the move towards GPUs was necessary, which changed the constraints for code performance. The ECP project catalyzed this transition by developing example codes that are sufficiently complex to be applied to problems of practical interest in various disciplines \cite{AMReX_JOSS, warpX, exaWind, musser2022mfix}. 

One of the products of these efforts is adaptive mesh refinement (AMR)-based solvers, which provide a new pathway for handling some of the emerging applications in propulsion and energy generation. Unlike conventional deflagration-based systems, such as gas turbines, new areas of interest include rotating detonation engines (RDEs)\cite{wolanski,gutmark_pecs,raman_arfm} and high-speed scramjet propulsion \cite{liu2020_review,shivank_pci,fureby2020_les}. In RDEs, a continuously propagating detonation wave traverses a confined flow path, typically orthogonal to the direction of propellant flow \cite{raman_arfm}. Due to the high velocity of the detonation wave, unsteady and incomplete mixing of the injected fuel and air results in secondary combustion processes, contributing to spurious losses. RDE flow paths are characterized by the intricate coupling of multiple physical processes. The presence of shocks and turbulence necessitates the use of stable numerical methods to minimize numerical dissipation. The strength of the detonation wave governs the stiffness of the chemical kinetics, which in turn influences the timescales of the reactions. While the dominant reaction zone is localized in the region near the shock, reactions occur throughout the flow field due to unsteady mixing. The smallest length scales in RDEs, comparable to the induction region length, can be as short as $\mathcal{O}(10)\mu$m for certain fuel mixtures, imposing stringent resolution requirements. These factors make direct numerical simulation (DNS) of RDEs computationally prohibitive. Simulations have generally struggled to resolve the shock-detonation structure with sufficient fidelity. Studies have demonstrated that the number of waves formed in the system is highly sensitive to numerical methods and grid resolution \cite{hayashi2021experimental}. Furthermore, the high computational cost restricts simulations to very short flow times, typically 3–5 ms, far short of the duration needed to achieve a true steady state. These challenges highlight the need for AMR-based solvers coupled with GPU-accelerated chemical kinetics computations\cite{raman_arfm, barwey2021neural}. Prior studies \cite{qian2020convergence, barwey2021neural} have shown that such approaches can enable the convergence of wave statistics, as AMR can effectively capture the reaction layer with adequate resolution. 

AMR uses a locally adaptive grid structure, which adjusts to the solution, in order to increase accuracy in specific parts of the domain \cite{dubey2014survey}. There are many approaches to AMR, including overset meshes \cite{kreiss1983construction,steger1987use}, octree-based refinement of local regions \cite{macneice2000paramesh, fryxell2000flash, grete2023parthenon}, and patch-based adaptation of solution pioneered by Berger and Oliger\cite{berger1984adaptive} and Berger and Colella \cite{berger1989local}, also referred to as the Berger-Oliger-Colella adaptive refinement\cite{colella2009chombo, hornung2002samrai, AMReX_JOSS}. There exists a wide range of open-source, general-purpose AMR frameworks and application codes that serve as foundational building blocks for developing domain-specific computational tools. This work focuses on leveraging the block-structured AMReX framework~\cite{AMReX_JOSS} to develop a compressible reacting flow solver with applications primarily, though not exclusively, in hypersonic propulsion. For a comprehensive overview of the existing block-structured AMR infrastructures, the reader is referred to the survey by Dubey et al. \cite{dubey2014survey}. The patch-based Adaptive Mesh Refinement (AMR) approach employed here constructs a hierarchy of grid patches at different AMR levels that lie on a coarse base mesh that spans the entire computational domain. Grid refinement strategies, guided by physics or geometry-based criteria, are used to tag cells for refinement. The refinement ratio, which may vary between successive AMR levels, determines the resolution of each refined region. The governing hyperbolic partial differential equations (PDEs) are solved across AMR levels, progressing from coarser to finer grids, followed by averaging down the finer-level solutions to the coarser levels. To ensure continuity and accuracy at patch and level boundaries, the framework employs ghost cells that facilitate the computation of solutions across boundary cells. This hierarchical approach enables efficient and scalable resolution of complex multi-scale phenomena in both space and time. The subsequent sections detail the various numerical approaches employed for spatial and temporal discretization, as well as the treatment of geometry and boundary conditions. This is followed by a series of canonical test cases used to evaluate the accuracy and robustness of the underlying numerical methods. Finally, the ability of the AMR-based solver to capture the disparate length scales inherent in scramjets is demonstrated through high-fidelity full-flow-path simulations of one such configuration. Additional applications address local microscopic instabilities that contribute to non-idealities in the macroscopic behavior of detonation-based propulsion systems.

Before proceeding to the description of the solution framework, we emphasize that the development of this solver was greatly aided by advances made within the ECP project, leading to new capabilities being made available to the user community on a regular basis. In particular, this solve utilizes the AMReX framework \cite{AMReX_JOSS}, but the PeleC solver \cite{peleC}, which has similar capabilities and has been the basis for the development of other related solvers \cite{peleLMeX, PeleSoftware}, was extensively used to identify approaches, understand the connection with the AMReX library, and the limitations of certain embedded boundary algorithms. Our priority is on high-speed and reacting flow applications, for which this new solver provides the basis. In accompanying publications, we have extended the capability to multiphase methods \cite{shivank_aiaa_2025,ral_thesis}, neural network-enabled chemistry for rapid GPU execution \cite{barwey2021neural,ullman2024_strat}, and for in situ learning of models \cite{vansh_superresolution,vansh_feedback}, and reduced order models for shock-containing flows \cite{yusuf_aiaa_2025, gibson_aiaa_2025}.

\section{Numerical approaches}

\subsection{Governing equations}
The code considers the formulation of the Navier–Stokes equations based on our prior work \cite{bielawski2023highly} for compressible turbulent shock-containing reacting flows relevant to scramjet and detonation engines. Mass transport is described by the following continuity equation
\begin{equation}
 \label{eq.mass}
    \frac{\partial \rho}{\partial t} + \frac{\partial (\rho u_i)}{\partial x_i} = 0,
\end{equation}
where $\rho$ is the density of the fluid, $t$ is time, $x_i$ and $u_i$ represent the spatial coordinate and the velocity component in the \textit{i}$^{th}$  direction. The momentum transport is given by
\begin{equation}
    \label{eq.mom}
    \frac{\partial \rho u_i}{\partial t} + \frac{\partial \rho u_i u_j}{\partial x_j} = - \frac{\partial p}{\partial x_i} + \frac{ \partial \tau_{ij}}{\partial x_j},
\end{equation}
where $p$ is the fluid pressure and $\tau_{ij}$ is the viscous stress tensor given by
\begin{equation}
    \label{eq.vis_tensor}
    \tau_{ij} = -\frac{2}{3} \mu \frac{\partial u_k}{\partial x_k} \delta_{ij} + \mu \bigg ( \frac{\partial u_j}{\partial x_i} + \frac{\partial u_i}{ \partial x_j} \bigg ),
\end{equation}
where $\mu$ is the dynamic viscosity evaluated from a tabulated temperature fit over the solution of collision integrals \cite{goodwin2017cantera}. The total chemical energy transport is governed via the following equation:
\begin{equation}
    \label{eq.chem_trans}
    e = \int_{T_0}^T C_P dT - \frac{p}{\rho} + \sum\limits_{k=1}^{N_s} \Delta h^0_{f,k} Y_k + \frac{1}{2} u_i u_i.
\end{equation}
In Eq. \ref{eq.chem_trans}, $T_0$ is the reference temperature (typically 298.15 K), $T$ and $C_p$ are the mixture temperature and the mixture specific heat, $Y_k$ and $\Delta h^0_{f,k}$ are the mass fraction and enthalpy of formation for the \textit{k}th species, and $N_s$ is the total number of species in the mixture.  The transport of the total chemical energy $e$ is given by:
\begin{equation}
\label{trans.e}
\frac{\partial \rho e}{\partial t} + \frac{\partial u_j \rho h }{\partial x_j}    = \frac{\partial }{ \partial x_j} \alpha \frac{\partial T}{\partial x_j} + \frac{\partial (u_i \tau_{ij} )}{\partial x_j} + \sum _{k=1}^N \left( h_k  \frac{\partial}{\partial x_j} \left( \rho  D \frac{\partial Y_k}{\partial x_j} \right) \right), \quad k \in\left[1, N_s\right],
\end{equation}
where $h = e + p/\rho$ is the total enthalpy of the mixture (sensible, chemical, and kinetic) and $\alpha$ is the mixture's thermal conductivity. The specific heat capacity, enthalpy, and specific heat ratio are all computed using NASA polynomials for thermodynamic data \cite{mcbride2002nasa}. Note that the specific heat and specific heat ratio are held constant below the minimum bound of the NASA polynomials, typically 300 K. The enthalpy and internal energy are computed relative to this minimum bounding temperature using the fixed specific heat. The gas is treated as an ideal gas and the governing equations are thus closed using an ideal gas equation of state. The viscosity and thermal conductivity for each species are fit as a function of temperature and pressure to reduce computational cost \cite{goodwin2017cantera}. Finally, the transport equation for a species $Y_k$ is given by:
\begin{equation}
\label{trans.yk}
\frac{\partial \rho Y_k}{\partial t} + \frac{\partial u_j \rho Y_k }{\partial x_j}    = \frac{\partial }{ \partial x_j} \rho D \frac{\partial Y_k}{\partial x_j} + \Omega_k,
\end{equation}
where $D$ is the spatially varying diffusivity of the mixture and is assumed equal for all species. The chemical source term (or the species production term) $\Omega_k$ for the \textit{k}th species is given by
\begin{equation}
\label{eq.omegaK}
    \Omega_k = \sum_{l=1}^{N_R} M_k (\nu_{k,l}^{\prime \prime} -\nu_{k,l}^{\prime} ) (\Pi_{f,l} - \Pi_{r,l}),
\end{equation}
where the stoichiometric coefficients of the reactants and products are given by $\nu^\prime_{k,l}$ and $\nu^{\prime \prime}_{k,l}$ for reaction $l$ and species $k$, and the forward
and reverse reaction rates are given by $\Pi_{f,l}$ and $\Pi_{r,l}$ respectively. The
forward rate constant is computed by the Arrhenius expression, and the reverse rate constant is computed by the forward and equilibrium rate constant. The chemical source term $\Omega_k$ introduces numerical stiffness due to fast chemical timescales, making the chemistry solver subroutines computationally expensive. This work employs operator splitting, assuming chemical timescales are much shorter than flow timescales, enabling a frozen flow approximation during kinetics evaluation \cite{macnamara2016operator}.

\subsection{Discretization}
A finite volume formulation is used to discretize the governing equations. Denoting the conserved variables $\mathbf{U}$, fluxes $\mathbf{F}$, and source terms $\mathbf{S}$, the governing equations can be formulated in a generalized form as 

\begin{equation}\label{eq:fv_semi_equation}
    \frac{\partial \mathbf{U}}{\partial t} + \frac{\partial \mathbf{F}}{\partial x_j} = \frac{\partial \mathbf{S}}{\partial x_j},
\end{equation}

where $\mathbf{U} = [\rho , \rho u_j, \rho e, \rho Y_k]$ are the conserved variables, $\mathbf{F}$ is the Euler term, and $\mathbf{S}$ represents the source terms that include bulk viscous dissipation, heat diffusion, viscous shear dissipation, enthalpy diffusion, and species diffusion terms. The governing equations are then integrated over each cell, which is treated as the control volume. Applying the divergence theorem allows the time rate of change of the volume-averaged conserved state $\mathbf{U}$ to be related to the sum of the fluxes and source terms evaluated on the cell faces:

\begin{equation}
    \frac{\partial \mathbf{U}}{\partial t} + \sum_j^{N_f} \mathbf{F}_j A_j = \sum_j^{N_f} \mathbf{S}_j A_j,
\end{equation}

where $A_j$ is the face area of the $j$th face over total faces $N_f$ enclosing a cell volume. The finite volume update procedure then requires the following: (1) convert averaged conserved states to cell-centered primitive states (2) interpolate primitive cell-centered states to cell face centers (3) compute the average flux over each face (4) update the averaged conserved states with a time integrator. 

\subsubsection*{(2.2.1) Approximate cell-centered primitive variables $q_i$}
The cell-centered primitive variables are obtained by transforming the conserved variables $\mathbf{U}$ to $\mathbf{Q} = [\rho, u_j, p, \rho Y_k]$ using the ideal gas equation of state $p = \rho R T$. This approximation is second-order accurate as higher-order terms including the derivatives of the conserved variables as in \cite{mccorquodale2011} are omitted.

\subsubsection*{(2.2.2) Interpolate cell-centered primitive variables to cell faces}
Primitive interpolation is done independently for each scalar primitive variable using a second-order limited interpolation routine of the form:

\begin{align}
    \label{eq:interppos}
    q_{i+1/2,L} &= q_{i+1} + w_{+} (q_i - q_{i+1}), \\
    \label{eq:interpneg}
    q_{i+1/2,R} &= q_{i+1} + w_{-} (q_i - q_{i+1}),
\end{align}

where $q_i$ denotes the primitive located at the cell center, $q_{i+1/2,L}$ is located on the left side of the $i+1/2$ interface (between cells $i$ and $i+1/2$), and $q_{i+1/2,R}$ is located on the right side of the $i+1/2$ interface. The weights $w_+$ and $w_-$ define the interpolation scheme and equivalent weightings of $w_{\pm} = 0.5$ correspond to central interpolation. A limiting procedure is applied which prevents large gradients and enforces monotonicity and total variation diminishing (TVD) properties over the domain. Cell-centered gradients are computed for each primitive variable using a second-order central difference stencil. Next, the slope ratios $r_+$ and $r_-$ are then calculated by dividing the ratio of neighboring gradients. A flux limiter function $\Psi(r)$ (such as minmod, van Leer, superbee) \cite{berger2005} returns the scaling parameters $\psi_+ = \Psi(r_+)$ and $\psi_- = \Psi(r_-)$. Finally, the weights $w_+$ and $w_-$ are computed by:

\begin{align}
    w_+ &= 0.5 \cdot \psi_+ + (1 - \psi_+) \label{eq:wpos} \\
    w_- &= 0.5 \cdot \psi_- \label{eq:wneg}
\end{align}

\subsubsection*{(2.2.3) Average flux across a face}
The spatial discretization of governing equations is split into viscous and Euler terms. The viscous terms utilize a second-order central scheme. The interface Euler flux $F$ in Eq.\ref{eq:fv_semi_equation} is obtained using a popular class of methods known as approximate Riemann solvers. For further details, the reader is directed to the work of Toro et al.~\cite{toro2013riemann}. In our discussion, we provide a brief overview of two approximate Riemann solvers: the Harten-Lax-van Leer contact scheme (HLLC)~\cite{toro1994hllc} and its variant, the HLLC-LM scheme~\cite{hllcLM2020shock}.

\paragraph*{(i) HLLC scheme} \mbox{} \\
The HLLC scheme effectively captures both contact and shear waves by introducing an additional contact wave with speed $S_*$ into the original two-wave HLL framework, making it a simple but accurate Riemann solver. This scheme also maintains entropy satisfaction and positive conservation, contingent on proper selection of wave speed. The HLLC interface flux $\mathbf{F}^{HLLC}$ can be written as:

\begin{equation}\label{eq:hllcF}
\mathbf{F}^{H L L C}= \begin{cases}\mathbf{F}_L & \text { if } S_L \geq 0, \\ \mathbf{F}_{* L}=\mathbf{F}_L+S_L \cdot\left(\mathbf{U}_{* L}-\mathbf{U}_L\right) & \text { if } S_L<0 \cap S_* \geq 0, \\ \mathbf{F}_{* R}=\mathbf{F}_R+S_R \cdot\left(\mathbf{U}_{* R}-\mathbf{U}_R\right) & \text { if } S_R>0 \cap S_* \leq 0, \\ \mathbf{F}_R & \text { if } S_R \leq 0,\end{cases}
\end{equation}

where two intermediate states, $U_{*L}$ and $U_{*R}$, are separated by the contact wave and and are determined by

\begin{equation}
\boldsymbol{U}_{* K}=\frac{S_K-u_K}{S_K-S_*}\left(\begin{array}{c}
\rho_K \\
\rho_K S_* \\
\rho_K v_K \\
\rho_K w_K \\
E_K+\left(S_*-u_K\right)\left(\rho_K S_*+\frac{p_K}{S_K-u_K}\right).
\end{array}\right)
\end{equation}

Here, $K=L,R$ and $U_L$ , $U_R$ are the reconstructed left and right face states, respectively. Nonlinear signal speeds on the left and right faces are estimated~\cite{einfeldt1988godunov} using as:

$$
S_L=\min \left(u_L-c_L, \ \widetilde{u}-\widetilde{c}\right), \quad S_R=\max \left(u_R+c_R, \  \widetilde{u}+\widetilde{c}\right),
$$

where $\widetilde{u}$ and $\widetilde{c}$ are estimated from the Roe averaging procedure:

$$
\widetilde{a}=\frac{a_L \cdot \sqrt{\rho_L}+a_R \cdot \sqrt{\rho_R}}{\sqrt{\rho_L}+\sqrt{\rho_R}}.
$$

The contact wave speed ($S_*$) is computed by \cite{batten1997choice}:

\begin{equation}
\boldsymbol{S}_* = \frac{p_R - p_L +\rho_L u_L (S_L-u_L) -\rho_R u_R (S_R-u_R)}{\rho_L(S_L-u_L)-\rho_R(S_R-u_R)}.
\end{equation}

\paragraph*{(ii) HLLC-LM scheme}\label{sec:hllc-lm}\mbox{} \\
The HLLC flux is susceptible to the generation of spurious oscillations and hot spots in strong normal shocks known as carbuncle phenomena~\cite{Quirk1997, dumbser-2004, kemm2018heuristical, pandolfi2001numerical, robinet2000shock}. The carbuncle instability is related to the computation of fluxes in the direction tangential to a grid-aligned shock. The HLLC-LM introduces a smooth decay of the acoustic dissipation in the low Mach regime, thereby providing sufficient dissipation to limit instabilities. Based on the work by Fleischmann et al., the HLLC-LM flux can be written using the HLLC flux [Eq.~\ref{eq:hllcF}] as 

\begin{equation}
\mathbf{F}^{H L L C - L M}= \begin{cases}\mathbf{F}_L & \text { if } S_L \geq 0, \\ \mathbf{F}_R & \text { if } S_R \leq 0, \\ \mathbf{F}_* & \text { else }\end{cases}
\end{equation}
with
$$
\mathbf{F}_*=\frac{1}{2}\left(\mathbf{F}_L+\mathbf{F}_R\right) \frac{1}{2}\left[\phi \cdot S_L\left(\mathbf{U}_{* L}-\mathbf{U}_L\right) + \left|S_*\right|\left(\mathbf{U}_{* L}-\mathbf{U}_{* R}\right) + \phi \cdot S_R\left(\mathbf{U}_{* R}-\mathbf{U}_R\right)\right],
$$
here, the classical HLLC flux has been rewritten in a central formulation by averaging the $F_{*L}$ and $F_{*R}$ intermediate fluxes with an activation function $\boldsymbol{\phi}$. The function, shown below, oscillates between 0 and 1, modulating the balance between the advection and acoustic dissipation components~\cite{hllcLM2020shock},
$$
\phi=\sin \left(\min \left(1, \frac{M a_{\text {local }}}{M a_{\text {limit }}}\right) \cdot \frac{\pi}{2}\right), \text{  with  \  }  M a_{\text {local }}=\max \left(\left|\frac{u_L}{c_L}\right|,\left|\frac{u_R}{c_R}\right|\right)
$$
and $M a_{\text {limit }} = 0.1$ for inviscid cases. For viscous cases, a careful selection of $M a_{\text {limit }}$ is required, as an aggressive value can affect the thickness of the boundary layer. The variable $M a_{\text{limit}}$ serves as a threshold, modulating the acoustic dissipation of the HLLC-LM scheme. This variable adjusts dissipation based on local velocities rather than the speed of sound for low Mach numbers, whereas the dissipation from advection terms remains constant. Based on numerical experiments with multiple viscous cases containing a strong shock, we found that a value of $M a_{\text {limit }} \in [0.01 \sim 0.02]$ is appropriate for our calculations. For further discussion, the reader is directed to the Appendix~\ref{appendix:hllclm}.

\paragraph*{(iii) Piro-Central Scheme}\mbox{} \\
An additional quasi-skew-symmetric central scheme (Piro-Central) is included to calculate the interface fluxes $F$ based on the work of Pirozzoli et al.~\cite{piro2017low}. The interface convective flux is defined as 
$$
\mathbf{F}_{face} = \frac{1}{8}\left(\rho_L+\rho_R\right)\left(u_{n L}+u_{n R}\right)\left(\boldsymbol{\varphi}_L+\boldsymbol{\varphi}_R\right),
$$
where $\boldsymbol{\varphi}=\left(1, u_i, E+\frac{p}{\rho}\right)^T$ for mass, momentum and energy fluxes, and the pressure flux is evaluated using the standard central interpolation. This scheme can be integrated with a locally activated diffusive flux scheme to form a hybrid scheme that can be used to simulate shock-containing flows with a higher accuracy compared to both HLLC and HLLC-LM schemes alone. 

\subsubsection*{(2.2.4) Time integration}
For time integration, there exists a choice between the second or third-order TVD explicit Runge-Kutta (RK) method by Shu and Gottlieb \cite{gottlieb1998total} in the solver. The chemical source term $\Omega_k$ introduces numerical stiffness due to the presence of extremely fast timescales within the chemistry model. Consequently, solver subroutines that account for chemical reactions via this source term are computationally intensive, and their efficient handling is critical to maintaining overall simulation tractability. In this study, an operator splitting strategy~\cite{macnamara2016operator} is adopted. The chemical kinetics are evaluated under the assumption of a frozen flow condition, as chemical timescales are significantly shorter than flow timescales. The solver is coupled with the Cantera~\cite{goodwin2017cantera} chemistry interpreter to compute the net production rates of species. To accelerate chemical source terms computation on GPUs, the solver is also integrated with the ChemGPU library \cite{barwey2021neural}, which recasts PDEs as a neural network-inspired matrix formulation. For further details on the formulation, the reader is directed to~\cite{barwey2021neural}.

\subsection{Representation of complex geometry}
For complex engineering geometries, the mesh generation process is often time-consuming and expensive. Meshing geometries in unstructured grids is more complicated and computationally intensive than in Cartesian grids due to their irregular and adaptive nature. Embedded boundary Cartesian grids vastly simplify the grid generation process as they reduce the grid generation process to computing intersections of the geometry with the underlying nonaligned Cartesian mesh. As a result, the mesh generation process becomes fast and robust in handling geometries of arbitrary complexity. In addition, the meshes are computationally efficient: typically, most of the mesh is regular Cartesian cells, with no additional geometric metrics to store \cite{berger_2017}. The challenge lies with the cut cells, which are more irregular than unstructured meshes and impose a significant time step restriction for explicit time integration methods. The geometry treatment in the present work is divided into two steps. The first step is the construction of a functional representation of the complex geometry. The second step is mapping the continuous functional representation of geometry onto the Cartesian mesh on all AMR levels and storing this information in an AMReX-distributed database class.

\subsubsection*{Functional representation of complex geometries}
The embedded boundary formulation for the treatment of complex geometries requires a mathematical implicit function that describes the relative position of the point $x$ (cell centers) with respect to the boundary of the geometry $\Omega$.  The function takes in the position $x$ as an input argument and returns a positive value at points inside $\Omega$, zero at the boundary of $\Omega$, and negative outside of $\Omega$. 
The geometries can be constructed in the following two ways.
\paragraph*{(i) Implicit functions}\mbox{} \\
AMReX provides several basic predefined implicit function classes that can be used to generate basic shapes such as spheres, cuboids, cylinders, planes, and ellipsoids. Constructive solid geometry (CSG) techniques can then be employed to generate more complex geometries by combining the implicit functions. To this end, the AMReX framework offers a variety of transformation operations, including intersection, union, complement, translation, rotation, lathe, and scale.
\paragraph*{(ii) CAD}\mbox{} \\
Engineering geometries such as scramjets and RDEs have complex, intricate shapes. Constructing these geometries using techniques like CSG can be challenging, especially when dealing with intricate fuel injectors, nozzle configurations, and their union with the combustor. CAD-based representation in the form of stereolithography (STL) files can be used to represent these geometries. AMReX provides a native database class to initialize the EB geometry information from the STL file. Similar to the implicit function approach, this requires the generation of a level set function from surface tessellations. The signed distance field needs to be generated for the finest level grid and requires the distance from the closest facet. This incurs large performance penalties as a brute force search through all facets is slow and inefficient. This cost then also scales with complicated and large STL files, which can often contain millions of facets. To avoid this bottleneck, the process is accelerated through the integration of EBGeometry framework\cite{Marskar2019, EBGeomGithub} with the solver. This framework embeds the facets into a bounding volume hierarchy (BVH) and uses a k-d tree approach for aggregating the facets with axis-aligned boundary boxes (AABB) to build a priority list for faster traversal through the surface tessellations. For more details on the algorithm, the reader is referred to~\cite{Marskar2019}.
\subsubsection*{Mesh generation}
The EB framework requires that for a given geometry embedded in a Cartesian grid, a volume mesh is produced that consists of the regular Cartesian cells away from the geometry and a list of irregular cells cut by the geometry, with the associated properties required by the finite volume solver. The functional representation of the geometry is passed to the AMReX framework through a public member function along with a geometry object that contains information about the domain and mesh on the finest level. The geometric information and a connectivity map are then initialized at the finest level by the AMReX foundation classes. The coarser levels are generated by coarsening the data on the finest levels to ensure that the volumes are consistent with the average fine-level volumes. The volume fraction (representing the fraction of volume of the cell inside the fluid region), area fractions (fraction of area of faces not covered by the solid body), and face and cell centroids are stored in a distributed database class. The cells are classified into three types based on the value of volume fraction: cells entirely covered by the body $\Lambda = 0$ are called covered cells, cells intersecting with the surface of the body are called cut cells $0 < \Lambda < 1$, and cells inside the fluid region $\Lambda = 1$ are called regular cells. The reader is referred to the AMReX documentation\cite{AMReX_JOSS} for more information on the embedded boundary implementation. 

\begin{figure}[!h]
\centering
      \includegraphics[width=.9\textwidth]{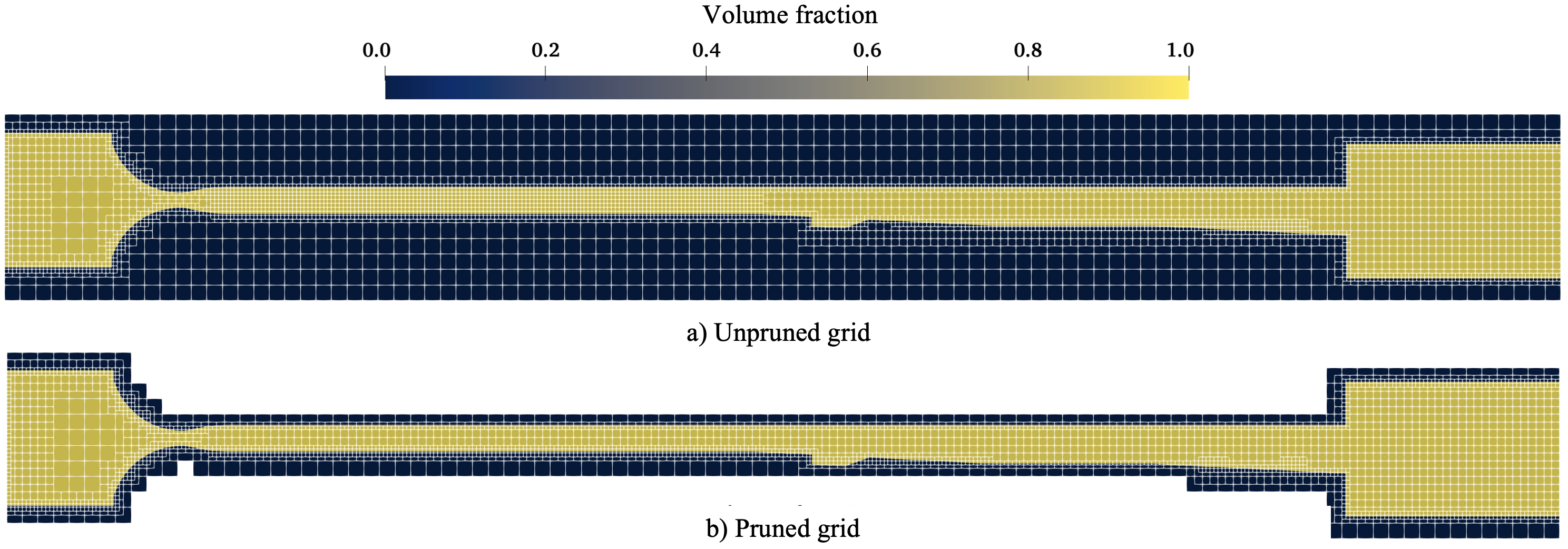}    
\caption{UVaSCF mesh with 3 AMR levels.}
\label{fig:UVa_pruned}
\end{figure}

\begin{figure}[!h]
\centering
      \includegraphics[width=.9\textwidth]{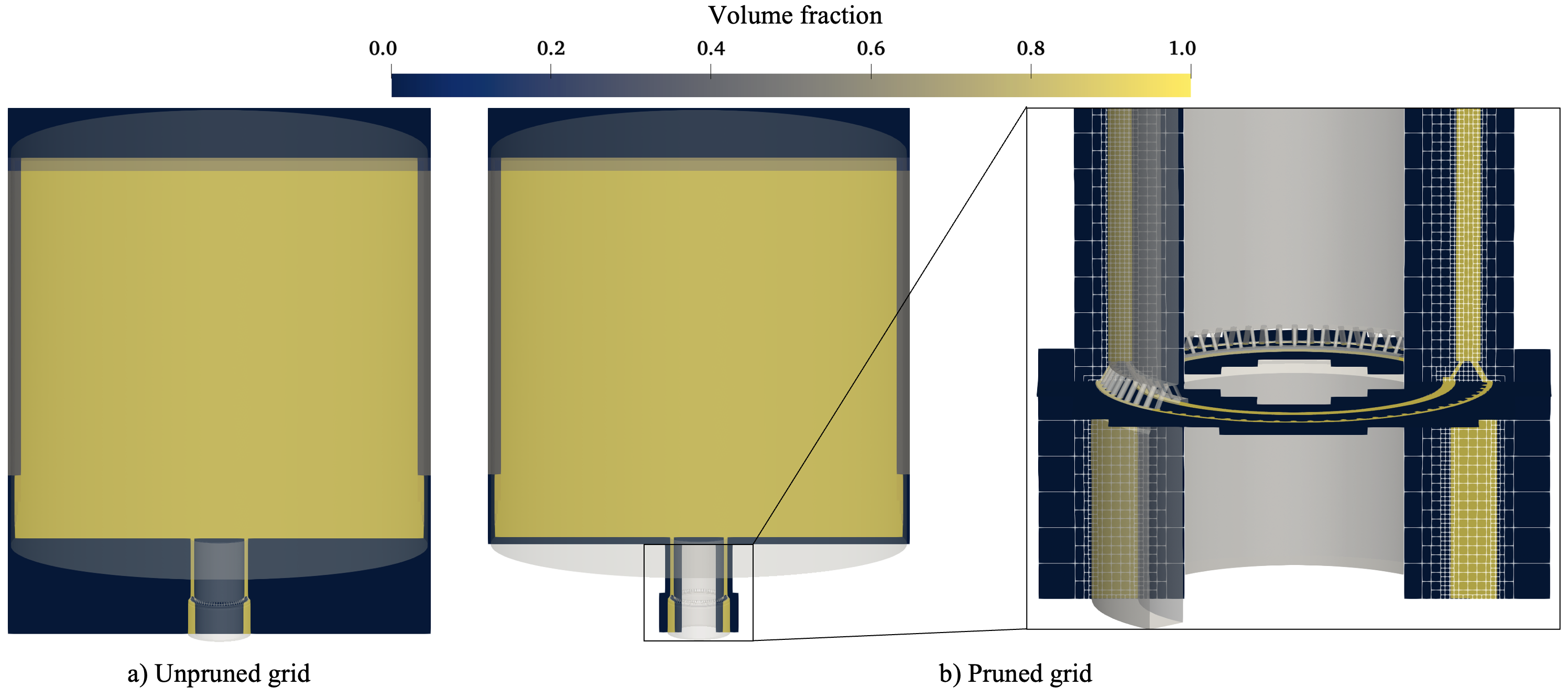}    
\caption{RDRE mesh with 5 AMR levels.}
\label{fig:RDE_pruned}
\end{figure}

For complex geometries such as RDEs and scramjets, there are often scenarios where the region enclosed by the EB mesh is often completely covered, or there is a requirement for the use of exhaust plenums, which then leads to the generation of a large region of covered cells outside the main device body as shown in Figs~\ref{fig:UVa_pruned} and \ref{fig:RDE_pruned}. The memory requirement increases for these geometries as these covered cells that are not used for computation are also stored. This motivates the use of mesh pruning, which removes fully covered grids at each AMR level. The pruning strategy is inspired by the MFIX-Exa code\cite{musser2022mfix}, where a second \texttt{BoxArray}  is generated by removing the covered grids from the original \texttt{BoxArray}. This leads to an overall reduction in memory that scales with finer mesh resolutions. A reduction of up to 40$\%$ in cell count was observed at higher AMR levels with the geometry shown in Fig.~\ref{fig:UVa_pruned}.
The code employs a load-balancing strategy that avoids computing convective, diffusive, and chemical source terms in covered cells. Thus, the pruning framework only helps to avoid the memory bottleneck for running large simulations with a reasonable number of compute nodes. 

\subsection{Embedded boundary method}
A cut-cell Embedded Boundary (EB) approach \cite{Pember1995,Colella2006} enables the treatment of arbitrary non-mesh conformal geometries. The embedded boundary will `cut' through cells arbitrarily, resulting in a new face on the interior of the cell and altered cell volume and centroid and face centroids and areas. AMReX provides the data structures representing the cut-cell geometry. The cut-cells require special EB treatment to enforce the desired boundary conditions. In particular, we must impose a new flux on the EB face, and the cut-face fluxes are modified according to their shifted centroid and area. This implementation interpolates the cut-face state to the shifted centroid before calculating the fluxes. Thus, it ensures that the wave speeds in the HLLC flux calculation use the face information at the corrected location.

While the EB implementation described above is sufficient to represent non-mesh conformal geometries, it suffers from a CFL constraint proportional to $1/\kappa_V$ where $\kappa_V$ is the volume fraction of the cut-cell. This CFL constraint is also known as the small cell problem and is prohibitive in complex geometries with cells where $\kappa_V \ll 1$. To alleviate the small cell problem, state redistribution \cite{SR1,SR2} is implemented in the solver through AMReX-Hydro \cite{HYDRO}. In state redistribution, cut-cells beneath a target threshold volume fraction redistribute the fluid state to neighboring cells in their merging neighborhood. The desired target volume fraction, with a default of 0.5, determines the merging neighborhood. This eliminates the dependence of the CFL number on the volume fraction. 

State redistribution has been successfully applied in compressible flows \cite{SR1,SR2}, but in the current form, it suffers from challenges of boundedness for the transport of passive scalars. This arises from distributing the merged neighborhood solution to the computational cells using a least squares limited reconstruction. The boundedness problem is thus analogous to the boundedness problem in flux evaluation in which face variables are reconstructed from cell values. Previous work addressing boundedness errors within flux evaluation includes \cite{larrouturou-specbound, subbareddy2017scalar}, which is applied to state redistribution as described below. The primary source of error in state redistribution is the limited least squares reconstruction of the conserved variables. Reconstructing mass, $\rho$, and species mass, $\rho Y_i$ separately results in $\sum \rho Y_i \ne \rho$, guaranteeing a non-physical species mass fraction. A simple second-order accurate solution is to reconstruct the mass, $\rho$, and mass fractions, $Y_i$, separately; this will give the sum of the mass fractions to be $1$. However, to maintain the boundedness of individual mass fractions, the smallest limiter among all species must be uniformly applied to the reconstruction of all the species mass fractions. These modifications to ensure species boundedness are implemented in the current solver.

\section{Canonical tests}
\subsection{1D Sod test}

A simple one-dimensional test for the solver is the Sod problem. This Riemann problem evolves over time to form several key wave features of compressible flow fields, including a rarefaction wave, a contact wave, and a shock wave. It is not a severe test condition, but it is useful for qualitatively demonstrating the abilities and limitations of the numerical scheme and for comparing numerical accuracy at different resolutions, including with adaptive mesh refinement. The initial conditions here have been scaled to conditions relevant to the application of the solver:

\begin{equation}
    0.0 \leq x \leq 0.5 : 
    \begin{cases}
        \rho_L = 1.0 \\
        U_L = 0.0 \\
        P_L = 101325 \\
    \end{cases}
    , \quad
    0.5 < x \leq 1.0 : 
    \begin{cases}
        \rho_R = 0.125 \\
        U_R = 0.0 \\
        P_R = 10132.5 \\
    \end{cases}
\end{equation}

\begin{figure}[!h]
\centering
      \includegraphics[width=.9\textwidth]{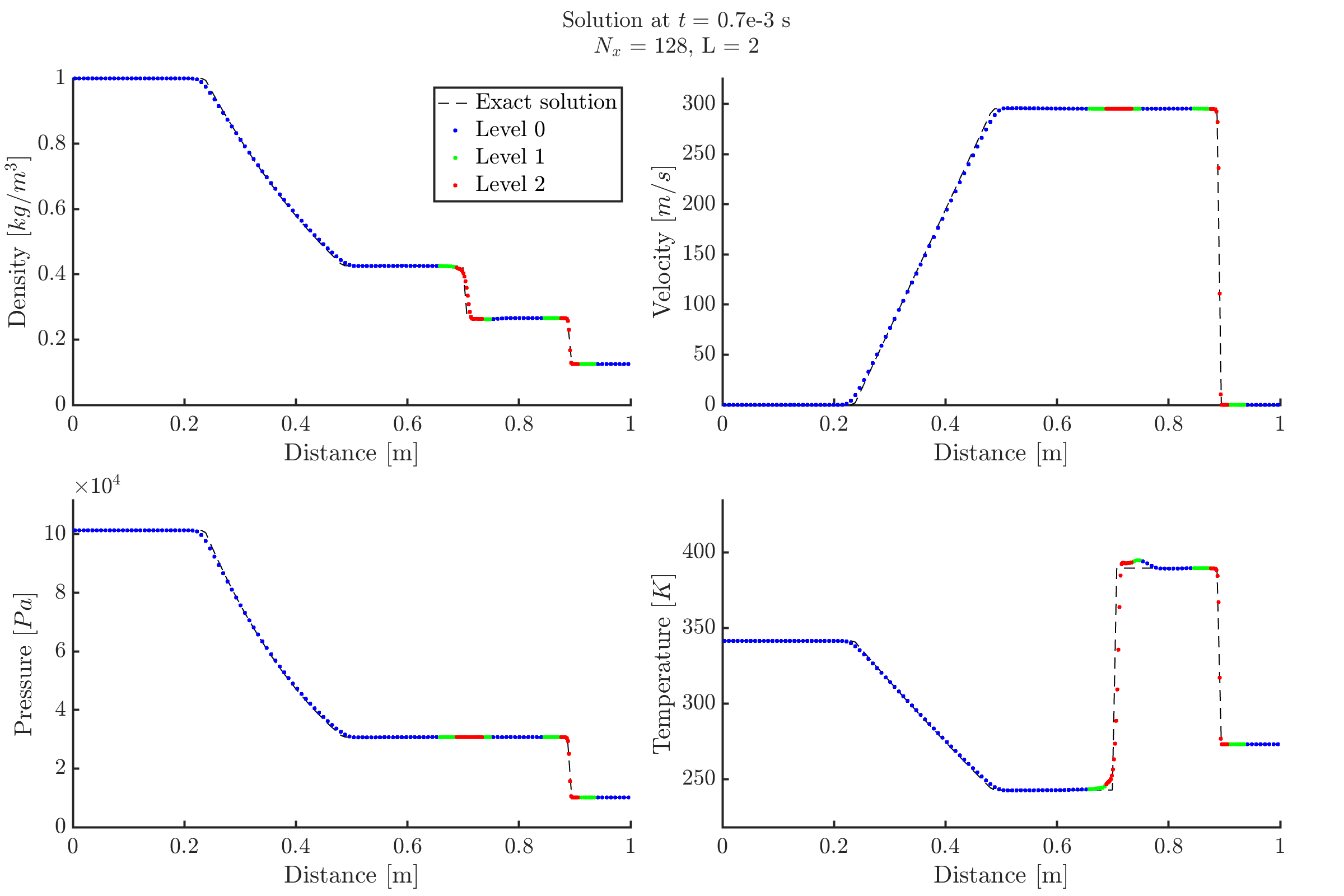} 
\caption{Scaled Sod problem with locally refined cell center values. This case has $N_x = 128$ cells and up to $L = 2$ AMR grid levels.}
\label{fig:Sod_128_L2}
\end{figure}

Here, the domain is $x \in [0, 1.0]$ and the final solution is advanced with CFL $= 0.3$ until time $t = $ 0.7e-3 s. Viscosity and reactions are not included. The species is $N_2$ and the thermodynamic properties such as the specific heat ratio $\gamma$ are not constant, but are a function of the temperature obtained from the Cantera chemistry package~\cite{goodwin2017cantera}. The interpolation procedure uses van Leer limiting with the HLLC flux function.  An exact Riemann solver~\cite{toro2013riemann} is used to generate the reference solution with a constant value $\gamma = 1.399$, which was evaluated to be a representative value in the solution temperature range. The property profiles for a case with $N_x = $ 128 cells on the base grid and up to $L = $ 2 levels of AMR levels are shown in Fig.\ref{fig:Sod_128_L2}. The grid refinement indicators were chosen to be the pressure gradient $>$ 5.0e5 and the density gradient $>$ 5.0, which are on the order of the expected property ranges in the domain and are intended to refine the steepest transient processes.

The solution is well captured for a cell count of $N_x = $ 128, with the only notable overprediction occurring in the temperature profile (corresponding to internal energy in the calorically perfect gas model). Cell-centered data points are colored by the AMR grid level, and there is a notable clustering of refinement near the contact wave and shock front. In general, an optimal value of the grid refinement indicators is not known before the simulation and there are several additional grid settings which affect the generation and location of local refinement (number of processors used, blocking factor, cells per grid, buffer layers, etc.). For a more complex test case, global gradient thresholds for indicators will not capture every wave of interest. In the particular case for $N_x = 128$ and $L = 0$, the grid refinement initially following the rarefaction wave has disappeared by the end of the simulation, although the refinement during the transient process will have altered the coarse grid solution upon averaging the finer grid data to the coarser grid.

\begin{figure}[!h]
\centering
      \includegraphics[width=.6\textwidth]{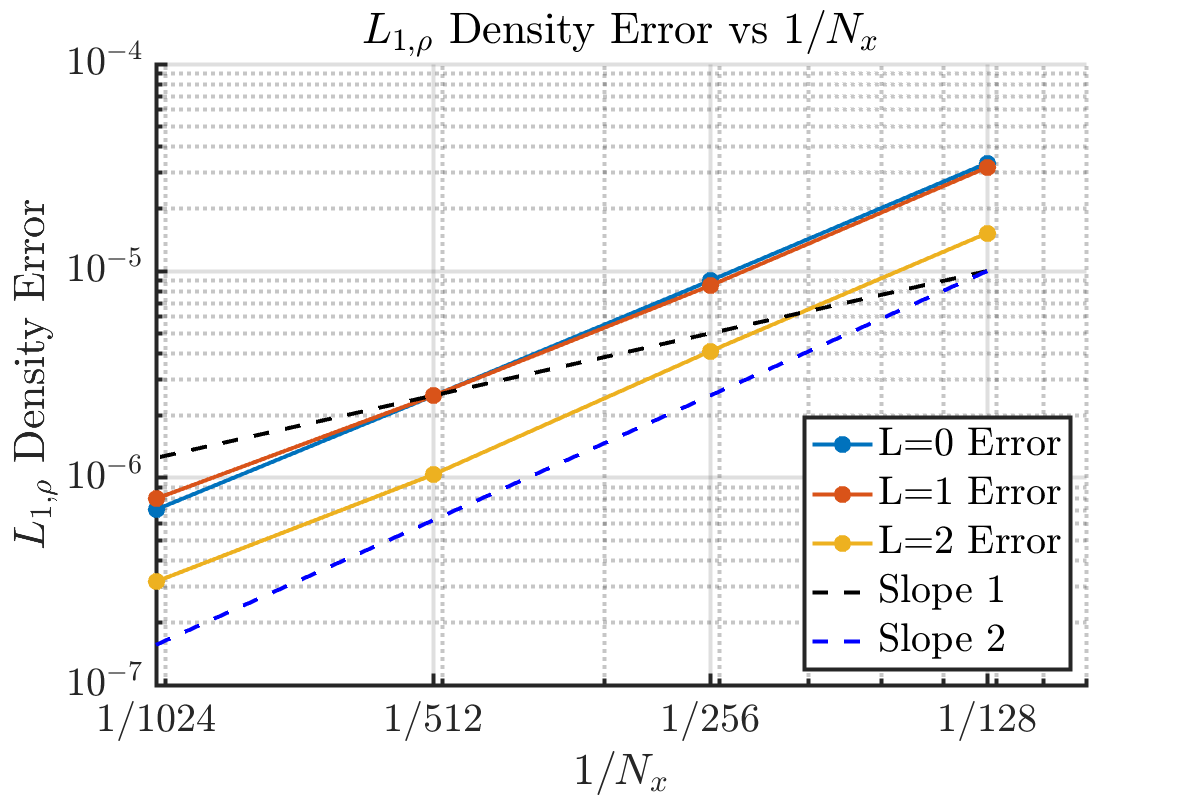} 
\caption{$L_{1,\rho}$ error norm for density in all cases.}
\label{fig:rho_convergence}
\end{figure}

The convergence of error for the $L_1$ error norm in the density provides insight into the accuracy obtained by using AMR levels. The simulations were performed using level 0 cell counts of $N_x = $ 128, 256, 512, and 1024 with up to $L = $ 0, 1, and 2 AMR levels. The cases with refinement levels are interpolated onto the corresponding level 0 grid size in order to compute the error in relation to the exact solution. Figure \ref{fig:rho_convergence} shows the convergence rates for the $L_{1,\rho}$ error norm for density in all cases. Interestingly, the cases with up to 1 AMR level are no more accurate than the cases with 0 AMR levels, as in Fig.\ref{fig:rho_convergence}. This demonstrates the challenge of selecting grid and refinement settings which contribute to a more accurate solution. When up to 2 AMR levels are used, the error decreases for all cases, although the convergence rate remains approximately the same compared to cases with fewer grid levels, as shown in Table \ref{tab:convergence_rates}. This behavior is expected because the underlying numerical schemes which dictate the convergence rate are identical and only the cell sizes have changed.

\begin{table}[h]
    \centering
    \begin{tabular}{ccccc}
        \toprule
        \multicolumn{2}{c}{Convergence rate for varying grid levels} \\ 
        \midrule
        Max grid level & Convergence rate \\
        \midrule
         $L = 0$ & 1.854 \\
         $L = 1$ & 1.7726 \\
         $L = 2$ & 1.865  \\
        \bottomrule
    \end{tabular}
    \caption{Convergence rates for $L_{\rho,1}$ density norm for cases with varying max grid levels.}
    \label{tab:convergence_rates}
\end{table}

\subsection{Homogeneous isotropic turbulence (HIT)}
Another insightful test case is that of homogeneous isotropic turbulence (HIT) in a periodic domain, which is suitable for investigating the dissipation properties of numerical methods. Note that AMR is not utilized in this case because the presence of volumetric features leads to mesh refinement across nearly all cells in the domain. HIT is simulated with different initial conditions in a cubic domain with edge length $L = 2 \pi$ and with varying interpolation limiting procedures: minmod, van Leer, superbee, and monotonized central~\cite{mignone2014}. The resolution is $64^3$ for all cases. The initial condition for HIT consists of a random solenoidal velocity field with energy spectrum:

\begin{equation*}
    E(k) = 16 \cdot u^2_{rms} \sqrt{\frac{2}{\pi}} \cdot \frac{k^4}{k_0^5} \exp \left (-2 \cdot \frac{k^2}{k_0^2} \right )    
\end{equation*}

where $k_0$ is the most energetic wave number, chosen to be 4, and:

\begin{equation*}
    u_{rms} = \sqrt{\frac{<u_iu_i>}{3}}.
\end{equation*}

Different conditions will be represented by the turbulent Mach number:

\begin{equation*}
    M_{t} = \frac{\sqrt{<u_iu_i>}}{<c>}    ,
\end{equation*}
 
 and Taylor-scale Reynolds number:

 \begin{equation*}
     Re_{\lambda} = \frac{<\rho> u_{rms}\lambda}{<\mu>}.
 \end{equation*}
 
The constant viscosity is obtained by:

\begin{equation*}
    \mu = \frac{<\rho> u_{rms} \lambda}{Re_{\lambda}}.
\end{equation*}

\begin{figure}[!h]
\centering
      \includegraphics[width=.40\textwidth]{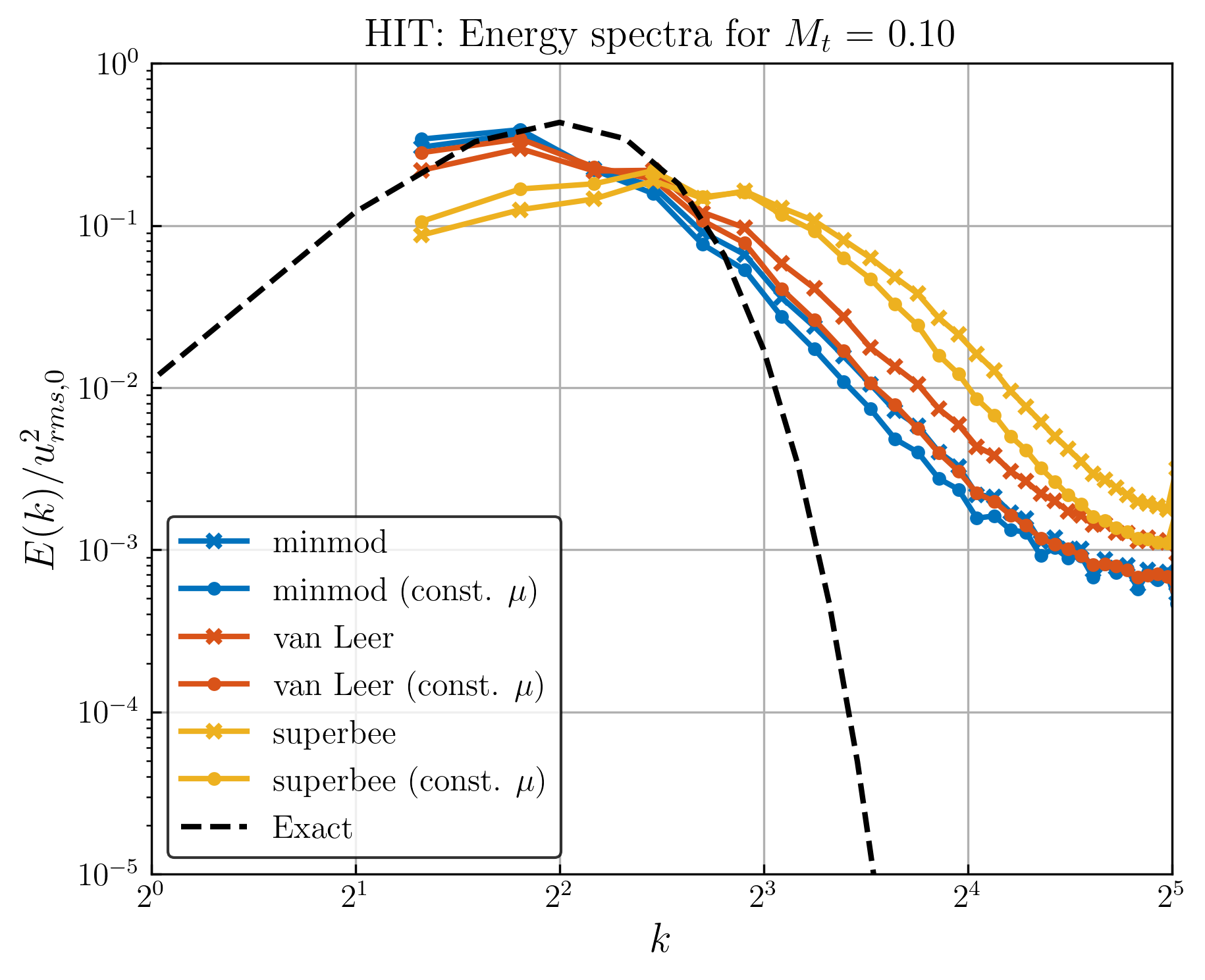}    
      \includegraphics[width=.40\textwidth]{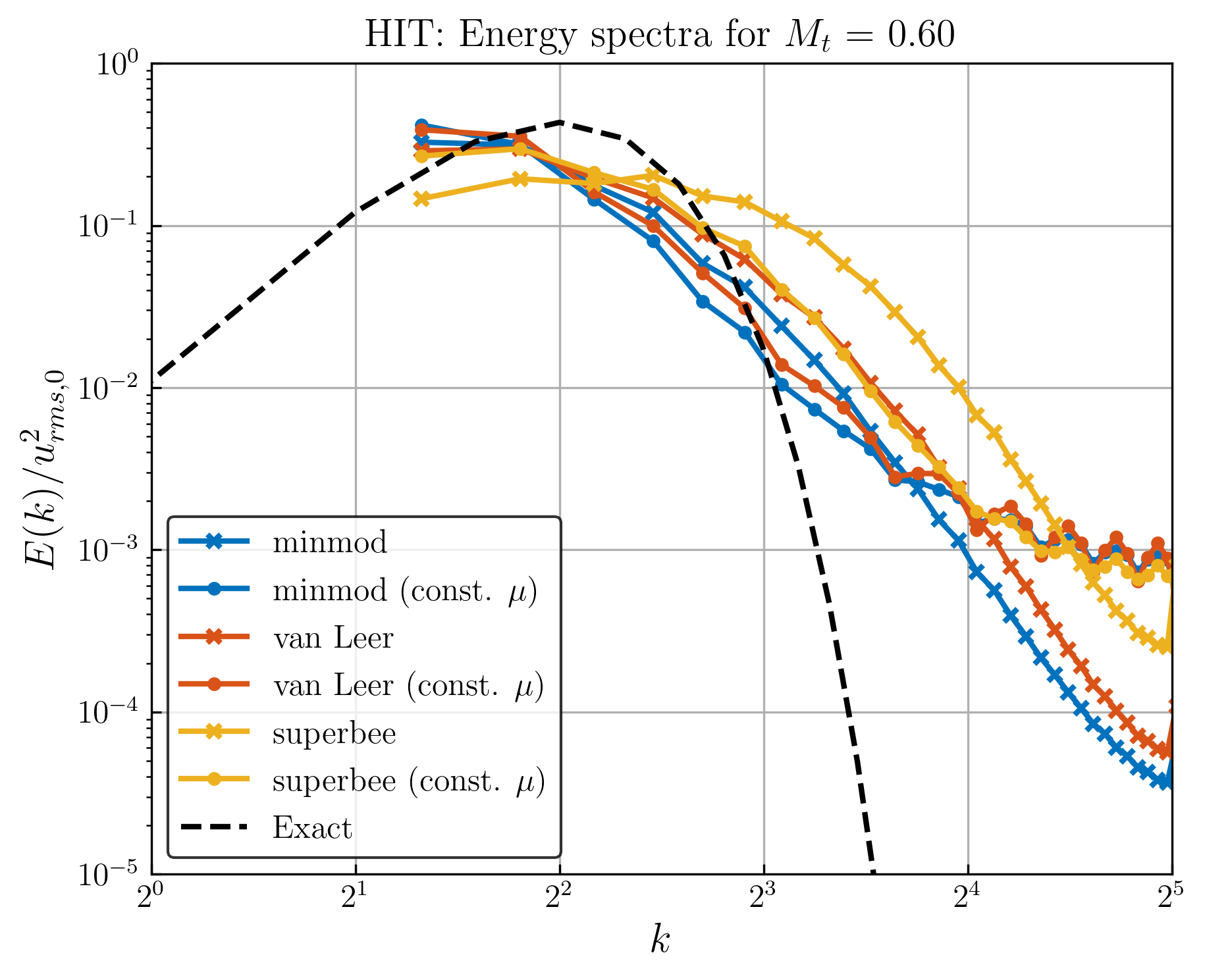}
\caption{Kinetic energy spectra over wave number for (left) $M_t = 0.1$ and (right) $M_t = 0.6$ cases.}
\label{fig:spectrumM01M06}
\end{figure}

Two conditions are simulated, one with turbulent Mach number $M_{t,0} = 0.1$ and another with $M_{t,0} = 0.6$, both with $Re_{\lambda,0} = 100$. In the second case with higher turbulent Mach number, the fluctuations at the initial condition are strong enough to generate weak shock waves called eddy shocklets \cite{johnsen2010assessment} which interact with and disrupt the turbulent velocity field throughout the simulation. The energy spectra for each condition can be used to understand the effects of the numerical schemes on the turbulent energy cascade, such as the amount of small-scale energy that can be maintained at a certain time in the simulation. Fig.\ref{fig:spectrumM01M06} shows the energy spectra at $t = 4 \tau$, in which the integrated area under the curve represents the total kinetic energy preserved over each wave number.

\begin{figure}[!h]
\centering
      \includegraphics[width=.39\textwidth]{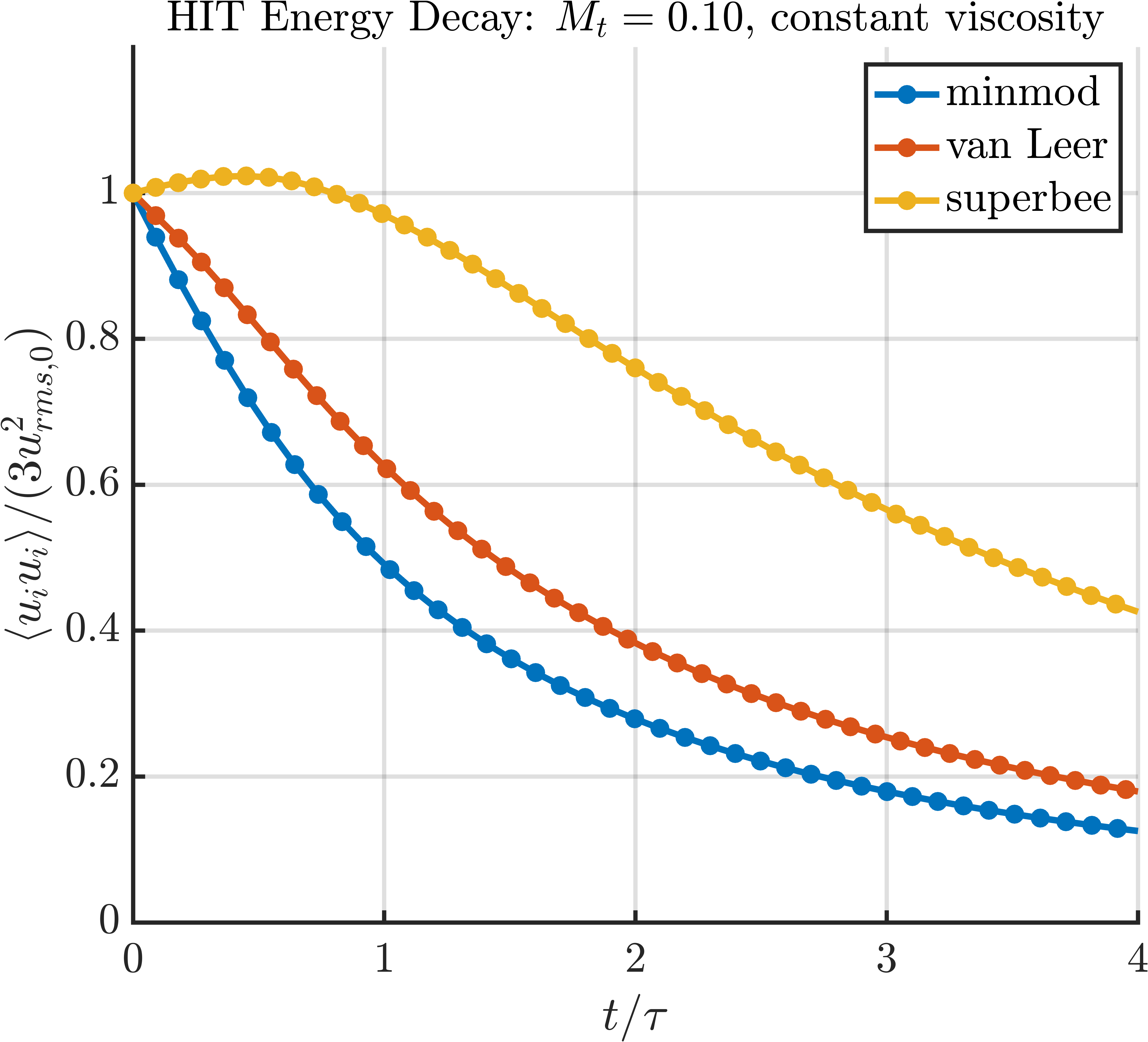}    
      \includegraphics[width=.39\textwidth]{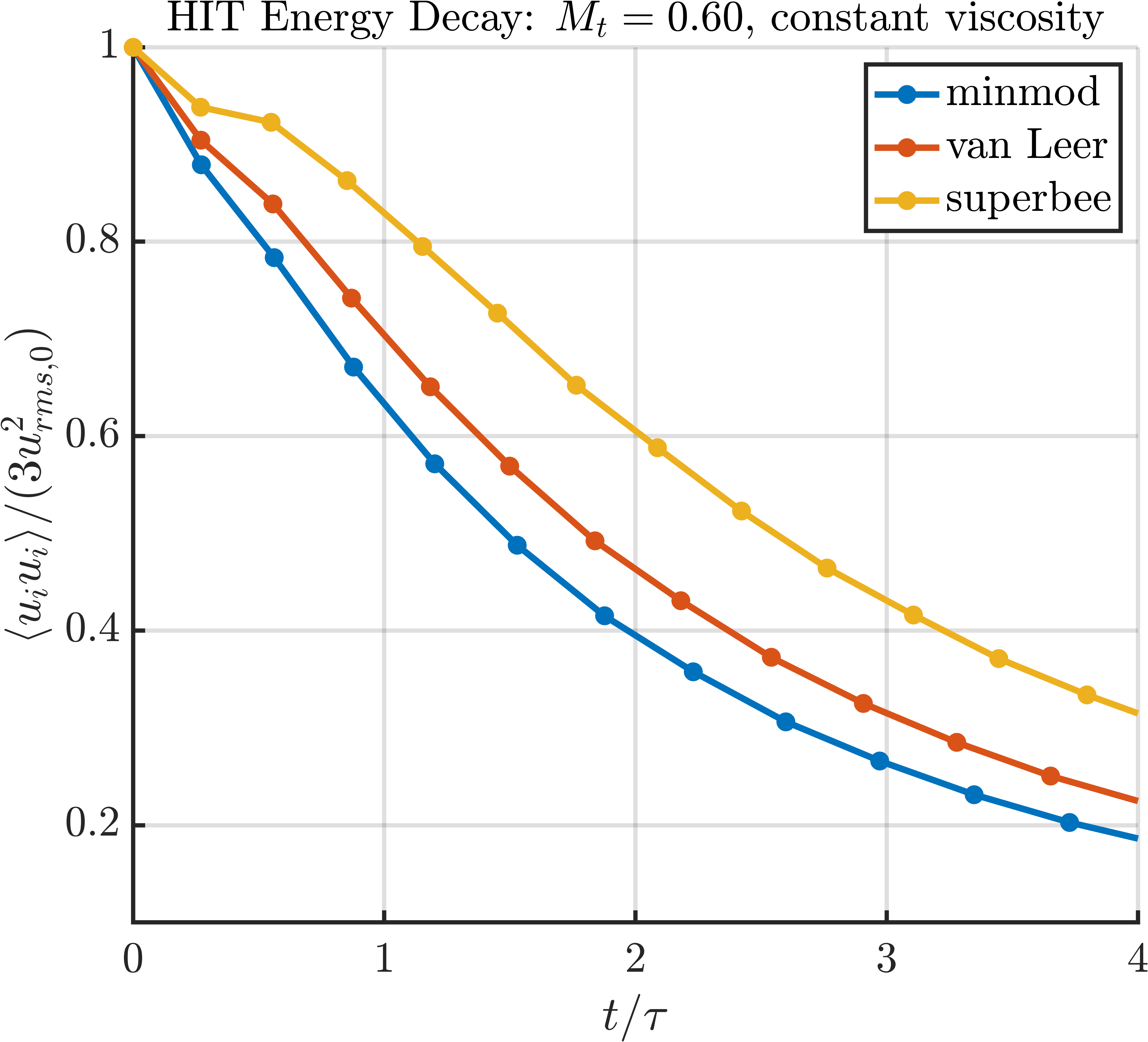}    
\caption{Kinetic energy over non-dimensionalized time for (left) $M_t = 0.1$ and (right) $M_t = 0.6$ cases.}
\label{fig.KE}
\end{figure}

Ideally, a solver will be able to conserve kinetic energy of the flow field while capturing both the turbulent fluctuations and the shocks, however in reality this is challenging as the numerical methods rely on dissipative numerical features to resolve the shock. This is observed in Fig.\ref{fig.KE}, which shows that the kinetic energy in the domain decays over non-dimensionalized time $\tau = \lambda / u_{rms,0}$ for the case of $M_t = 0.6$ and constant viscosity. The decay is in fact strongly dependent on the slope limiting procedure used, with minmod being most dissipative, van Leer having moderate dissipation, and superbee being least dissipative and in fact adding energy initially by increasing compressibility effects.

\subsection{Turbulent channel}
\begin{figure}[!h]
	\centering
	\begin{tabular}{ll}
		\makebox[0pt]{\raisebox{21ex}{\hspace{-1.3ex}{(a)}}} \hspace{0.2cm} \includegraphics[trim={0cm 0cm 0cm 0cm},clip,width=.45\textwidth]{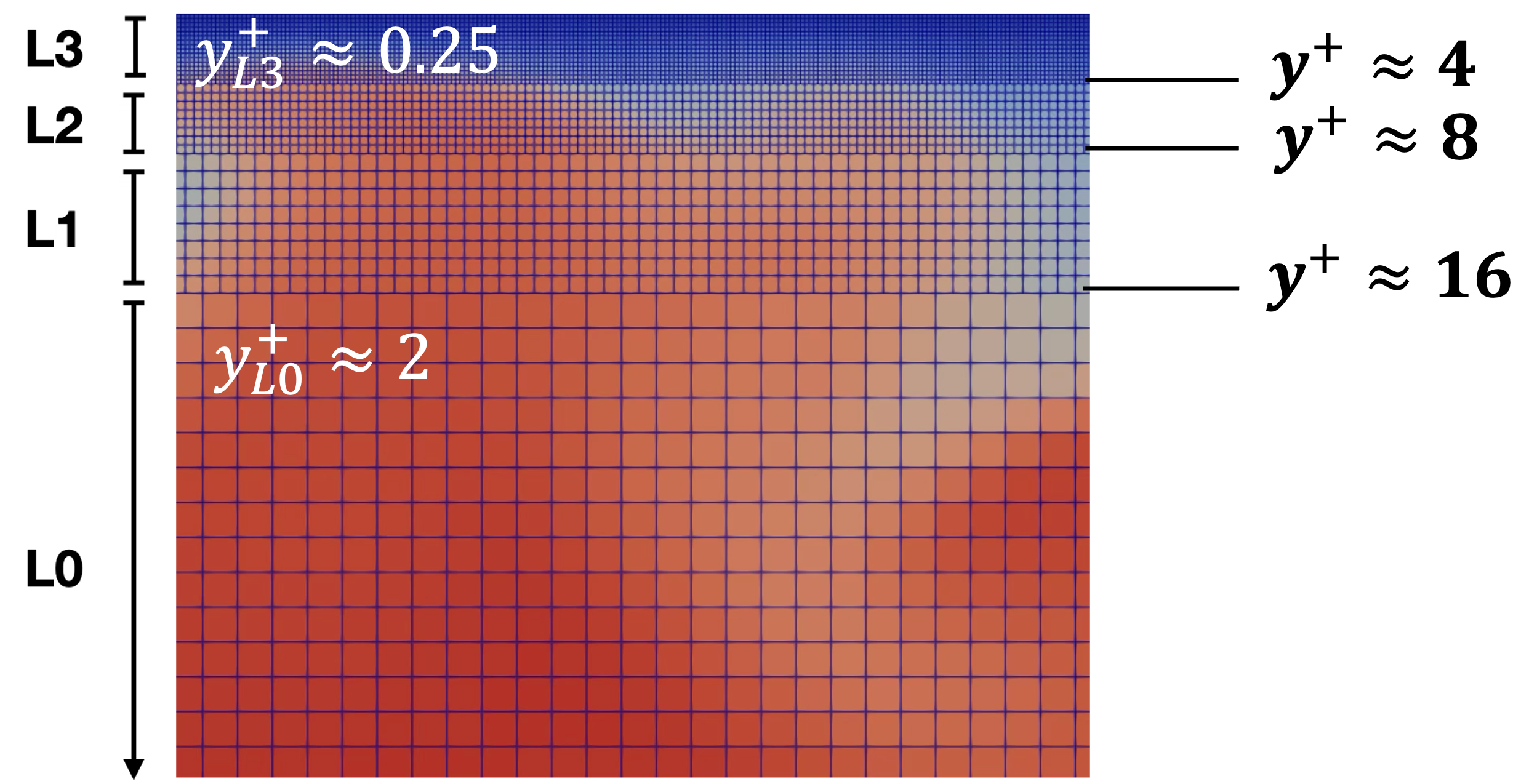}    & \makebox[0pt]{\raisebox{21ex}{\hspace{-1.3ex}{(b)}}} \hspace{0.2cm} \includegraphics[trim={20cm 10cm 1cm 8cm},clip,width=.6\textwidth]{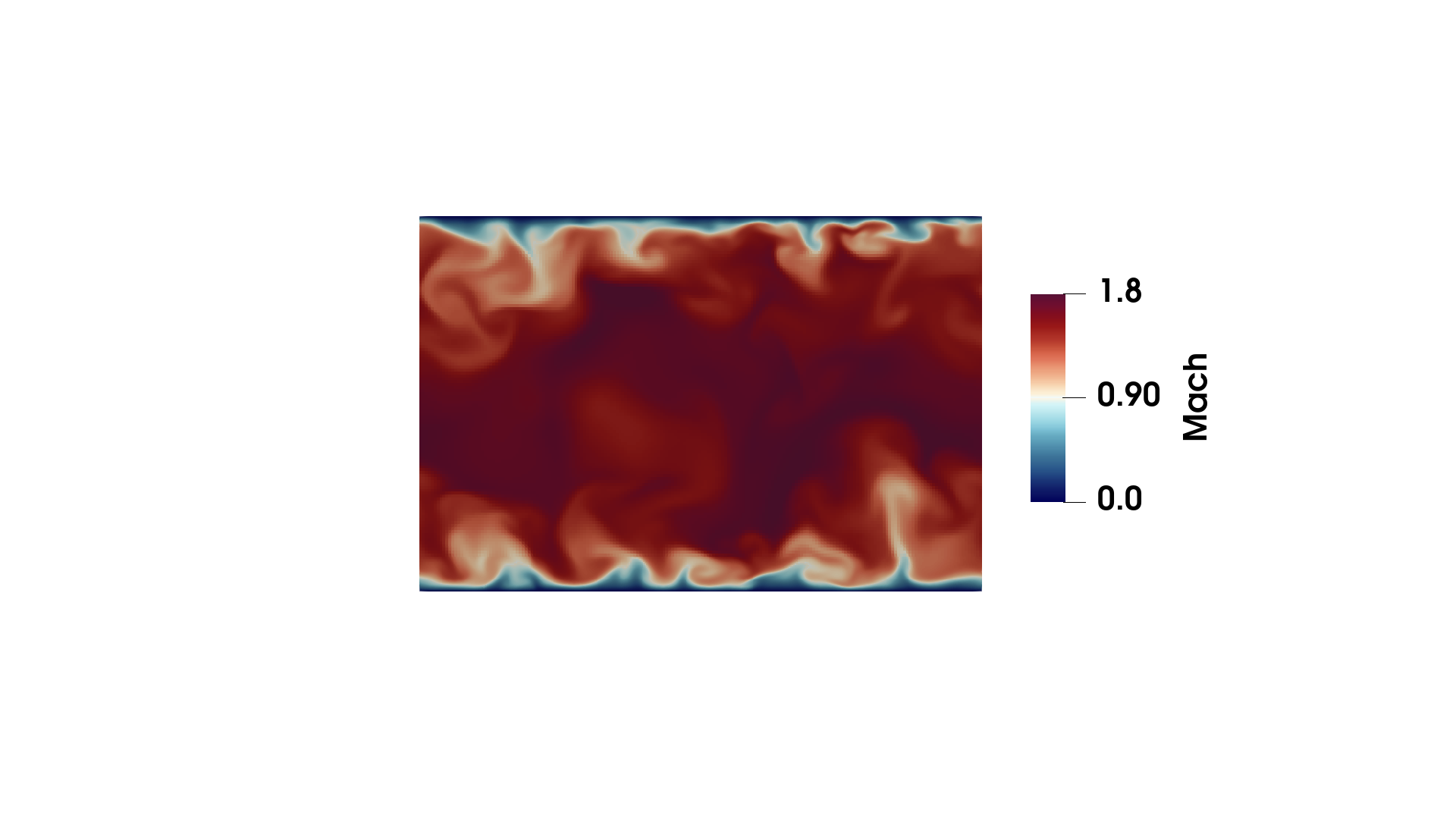}   
  \end{tabular}
	\caption{(a) Mesh refinement used in channel flow simulation (b) cross-sectional view of the instantaneous Mach number.}
	\label{fig.instant3}
\end{figure}

\begin{figure}[!hbt]
\centering
      \includegraphics[trim={0cm 0cm 0cm 1cm},clip,width=.7\textwidth]{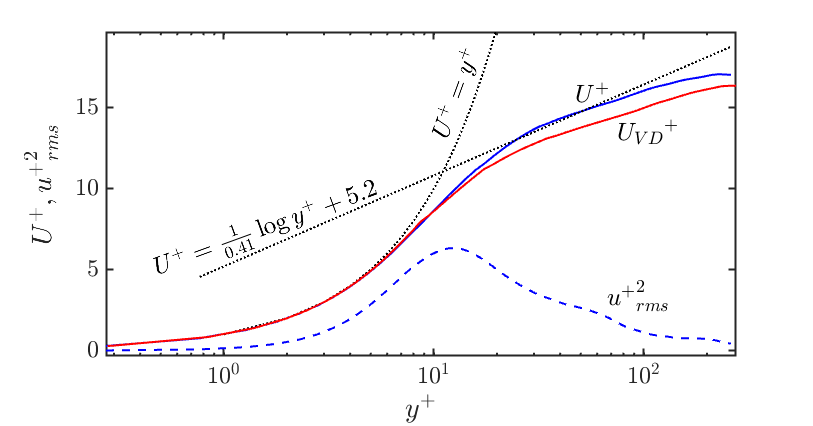}    
\caption{Time and spatially averaged profiles of mean axial velocity (solid lines) and axial turbulent intensity (dashed line) expressed in wall coordinates. Mean velocity profile following the Van Driest transformation is shown in red color. Dotted lines show the profiles of the law of wall \cite{Pope_2000}.}
\label{fig.channel_mean}
\end{figure}

Turbulent channel flow simulations are conducted at a centerline Mach number \( M_c = 1.8 \) and a friction Reynolds number \( Re_\tau \approx 260 \). The friction Reynolds number is defined as:
\begin{equation}
    Re_\tau = \frac{\rho_w u_\tau h}{\mu_w},
\end{equation}
where $\rho_w$ is the density at the wall, determined \emph{a posteriori} in the simulation, $\mu_w$ is the dynamic viscosity at the wall,$h$ is the channel half-height, and $u_\tau$ is the friction velocity. The friction velocity $u_\tau$ is given by
\begin{equation}
    u_\tau = \sqrt{\frac{h}{\rho_w} \frac{dp}{dx}},
\end{equation}

where $dp/dx$ represents the mean pressure gradient estimated using a turbulent friction factor correlation~\cite{Pope_2000}. The computational domain is a rectangular box with a height of \( 2h \), a length of \( 12h \), and a width of \( 3h \). Periodic boundary conditions are applied in the streamwise (\( x \)) and spanwise (\( z \)) directions, while the wall-normal (\( y \)) boundaries are modeled as no-slip, adiabatic walls. The simulation used four levels of mesh refinement, with three levels specifically refining the near-wall region. The final mesh had approximately 412 million cells, with a minimum and maximum resolution of 0.25 and 2 in wall units, respectively. The extents of different mesh refinement levels are shown in Fig.\ref{fig.instant3}(a), while Fig.\ref{fig.instant3}(b) displays the instantaneous Mach number in a cross-sectional plane. Temporally and spatially averaged profiles in wall units are presented in Fig.~\ref{fig.channel_mean}. The time-averaging was conducted over a period equivalent to 0.7 flow-through times, and spatial averaging was applied along the streamwise ( x ) and spanwise ( z ) directions. 

To express flow variables in wall units, a length scale of $(\mu_w/(\rho_w u_\tau))$ and a velocity scale $u_tau$ are used~\cite{Pope_2000}. The mean velocity profiles, transformed via the Van Driest transformation~\cite{huang1994van} which accounts for the compressibility, are also included in the plot. The transformed velocity, ${U^+}_{VD}$, is calculated as:
\begin{equation}
    U^+_{VD} = \int_0^{U^+} \sqrt{\rho/\rho_w}\, dU^+,
    \label{eq.vd}
\end{equation}
where $\rho$ is the time-averaged density that is a function of the distance from the wall. As expected, the mean axial velocity profiles follow the law of the wall.

\section{Applications}
\subsection{Scramjet}

\begin{figure}[!htb]
    \centering
    \begin{subfigure}{1.0\textwidth}
        \centering
        \includegraphics[width=.6\textwidth]{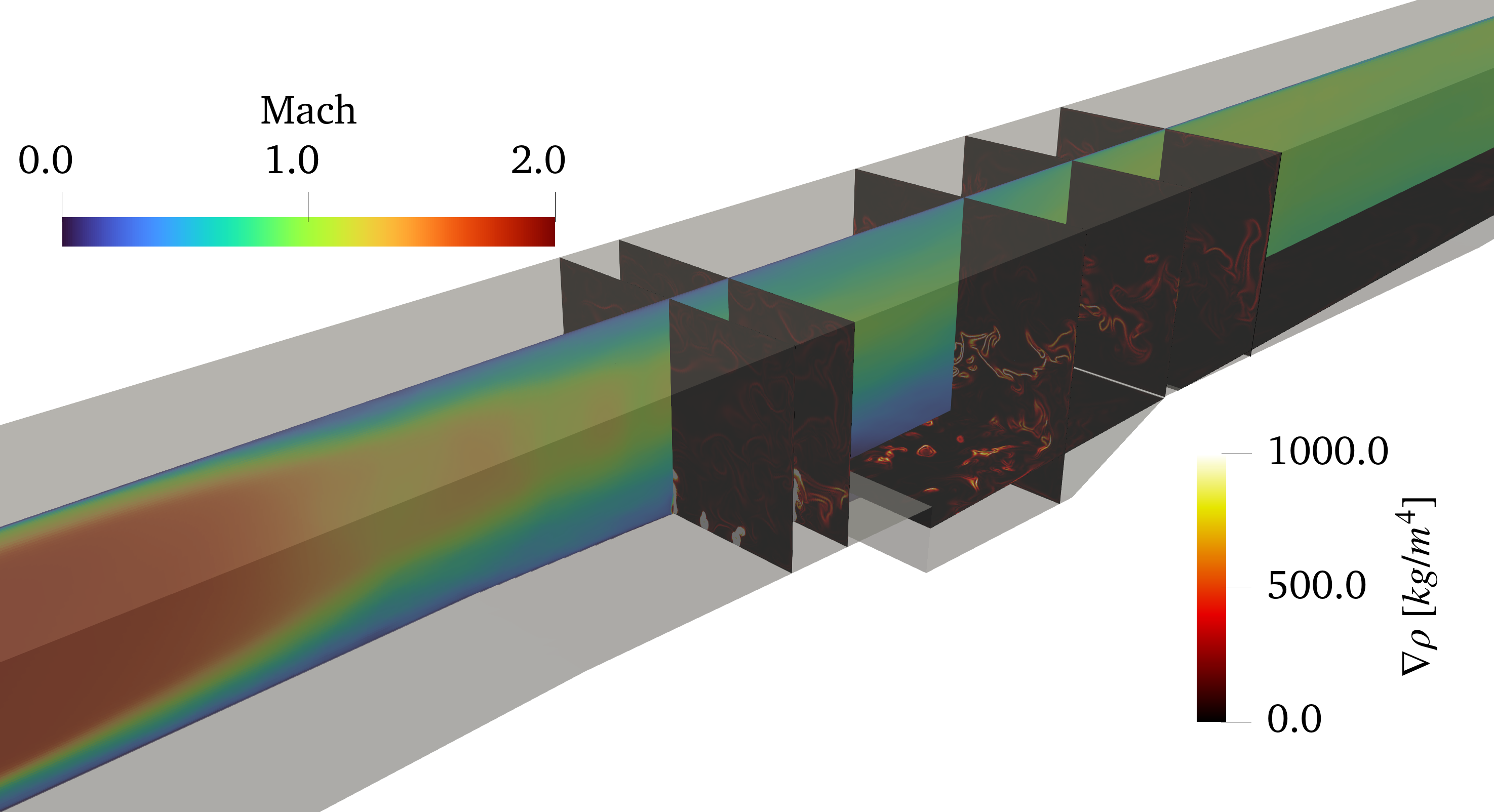} 
        \caption{Mean mach number on the centerline slice and density gradient on the spanwise and vertical slices.}
        \label{fig.scramjet.mach_rhograd}
    \end{subfigure} \\
    \begin{subfigure}{1.0\textwidth}
        \centering
        \includegraphics[width=.6\textwidth]{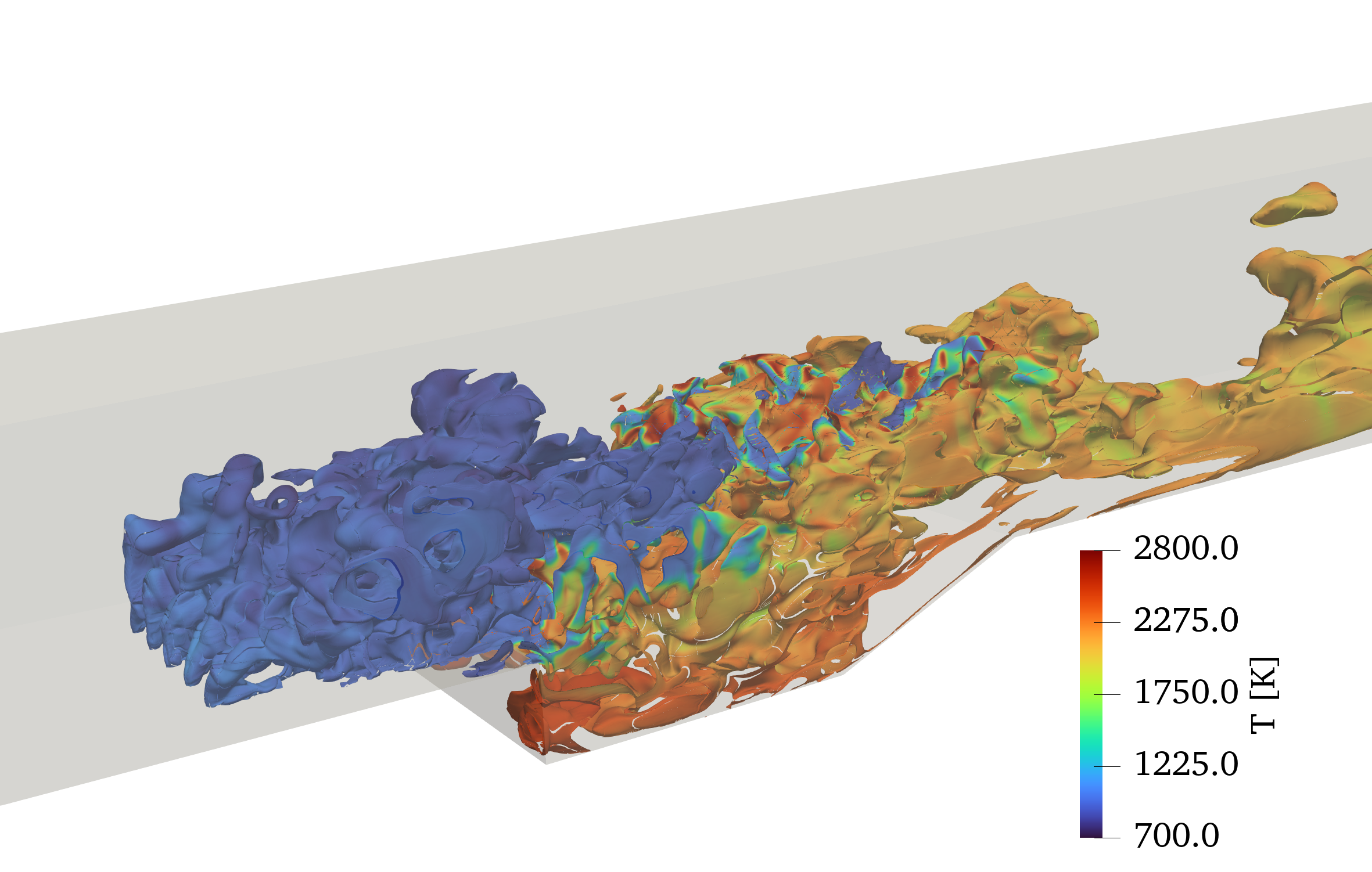} 
        \caption{Stoichiometric mixture fraction surface colored by temperature}
        \label{fig.scramjet.stoich}
    \end{subfigure}
    \caption{Results from a simulation of an experimental flowpath with a direct injection cavity flameholder in dual-mode ramjet/scramjet operation. The cavity step height is 9 mm.}
    \label{fig.scramjet}
\end{figure}

Full flowpath simulations of the University of Virginia Supersonic Combustion Facility (UVaSCF) with a cavity flameholder operating in dual ramjet/scramjet mode were carried out \cite{rauch-scramjet-scitech}. Simulation of the full flowpath is critical to capturing the correct fluid state and turbulence characteristics entering the combustor region. Conditions correspond to a flight Mach number of approximately 5 with a total temperature of 1200 K, resulting in a nozzle exit Mach number of 2 with a static temperature and pressure of 698 K and 38.7 kPa, respectively. Direct injection of gaseous ethylene (C$_2$H$_4$) occurs through two rows of five injectors, each with a diameter of 0.53 mm, resulting in a global equivalence ratio of $\Phi=0.32$. The scale separation of the largest, $O(10^0)$, and smallest geometric features, $O(10^{-3})$, is on the order of $O(10^3)$. Considering the smallest turbulence and flame scales, $O(10^{-6})$, a scale separation of over six decades becomes apparent. This highlights the need for the AMR-based approach for efficiently simulating the disparate scales.

A base resolution of 3.625 mm is used in combination with three AMR levels, resulting in a maximum resolution of 56 $\mu$m. The peak resolution provides ten cells across the injectors, eight cells across the flame thickness, and is approximately 2.5 times larger than the Kolmogorov scale. Targeting refinement on the heat-release allows the reaction layer to be fully resolved, this is further coupled with refinement of high temperature and density gradient regions to capture the coupling of compressibility, turbulence, and heat-release in the cavity region. Figure~\ref{fig.scramjet.mach_rhograd} shows the shock-train terminating near the fuel injectors, resulting in a largely subsonic flow over the cavity. High density gradients occurs from the choked injectors expanding into the core flow as well as over the cavity due to heat-release associated density changes. In Fig.~\ref{fig.scramjet.stoich} an isosurface of the stoichiometric mixture fraction ($Z=0.066$) is colored by temperature, this shows that peak temperatures are reached in the cavity and then convected downstream. The simulations agreed well with experimental relative pressure rises and reproduced a cavity stabilized flame with thermal choking. The reader is referred to \cite{rauch-scramjet-scitech} for further details.

\subsection{Oblique Detonation Waves}
\begin{figure}[!h]
    \centering
    \includegraphics[width=0.85\textwidth]{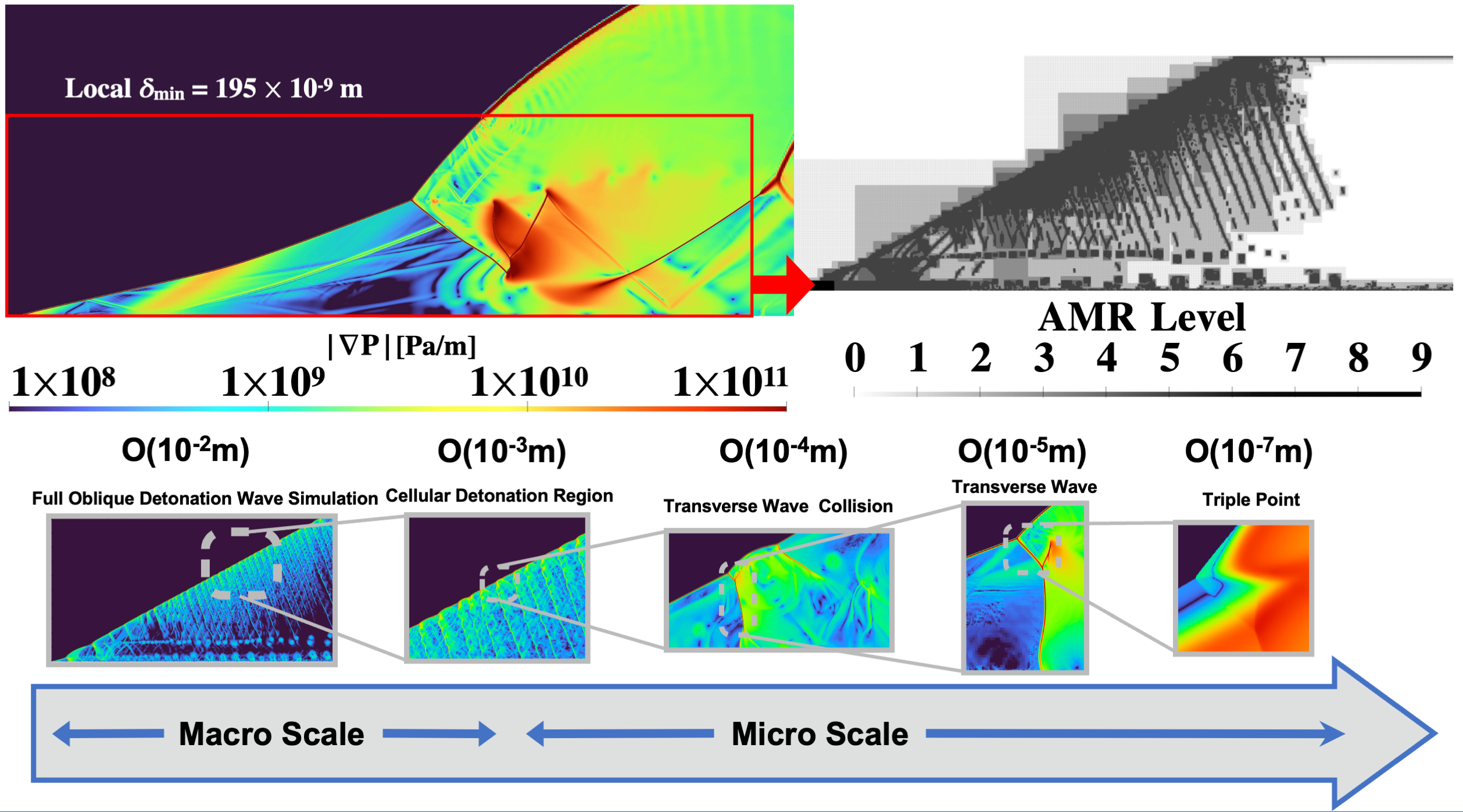} 
    \caption{Two-dimensional oblique detonation highlighting the range of physical scales present.}
    \label{fig:2DODW}
\end{figure}

Propulsion systems based on detonation have garnered growing attention because of their proposed higher thermodynamic efficiency, simpler design, and reduced weight \cite{wolanski}. There has long been interest in studying stationary detonation waves anchored to an oblique shock wave (OSW) as they are thought to offer performance comparable to ideal ramjet designs at much higher flight Mach numbers, indicating a possible means of extending the speed and range of hypersonic air-breathing vehicles utilizing oblique detonation wave engines (ODWEs) \cite{dunlap, wolanski, kailasanath}. Among the varying designs of ODWEs, some may involve the presence of conical detonation waves (CDWs) and oblique detonation waves (ODWs), which are detonation waves initiated by a conical or wedge geometry, respectively. To realize the usage of CDWs and ODWs in propulsion devices, its structure and stability must be further understood. With this in mind, a two-dimensional simulation of a wedge-induced oblique detonation is conducted under inflow conditions of $M = 5$, $T = 500 \ K$ and $p = 1  \ atm$ for a stoichiometric $H_2$-air reactant mixture. The dimensions of the domain are $L_x$ = 4 cm and $L_y$ = 1 cm. Triple-point instabilities along the detonation surface can be spatially refined through the application of adaptive mesh refinement. Figure \ref{fig:2DODW} illustrates the range of scales from the lengthscales of the detonation on the order of centimeters to the lengthscales of triple-points on the order of hundreds of nanometers. The finest resolution at the location of peak pressure gradients along the detonation front and transverse waves is 781 nm. This spatial resolution with 8 levels of refinement is performed in order to adequately capture the detonation induction length with a minimum of 40 cells per induction length broadly throughout the simulation domain.  Adaptive mesh refinement allows a 655 million cell case, if the finest resolution was applied uniformly throughout the domain, to be simulated with 50 million cells. Further details on this study can be found in Ref. \cite{abisleiman}.

\subsection{Detonations in Stratified Mixtures}

Practical rotating detonation engines (RDEs) and other related engine concepts typically mitigate flashback by using non-premixed fuel and oxidizer, which thereby requires the reactants to mix over short timescales in the combustion chamber \cite{raman_arfm}. Typically, this mixing process does not complete before the arrival of the next detonation wave, leaving the wave to process stratified reactant mixtures with localized fuel-rich and fuel-lean regions. To isolate the effects of mixture inhomogeneities on the propagation and structure of detonation waves, canonical channel detonation simulations were conducted using stratified reactants with varying length scales \cite{ullman2024_strat}. Using a digital filtering algorithm \cite{kempf_df}, the H$_2$-air equivalence ratio fields were initialized with prescribed bounds and integral length scales. These length scales were varied, while the bounds and ambient conditions ($p$ = 0.5 atm, $T$ = 297 K) were held constant. The equivalence ratio bounds were set such that the lean regions were pure air (i.e., $\phi$ = 0) and the rich regions were H$_2$-air at roughly $\phi$ = 1.3. The mean equivalence ratio was then roughly 0.65.

\begin{figure}[!h]
    \centering
    \includegraphics[width=\textwidth]{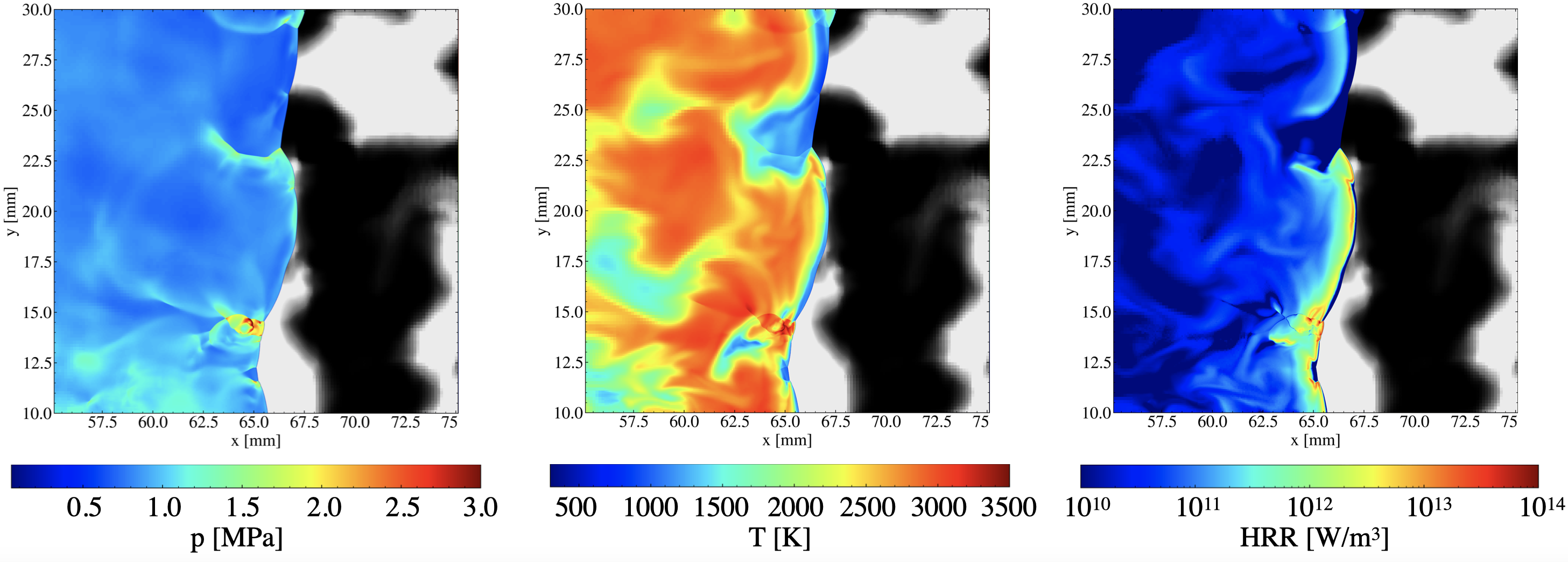} 
    \caption{Instantaneous pressure, temperature, and volumetric heat release rate fields along mid-plane of a 3-D stratified detonation case. Grayscale ahead of the wave denotes H$_2$ mass fraction, where black represents fuel-lean and white represents fuel-rich.}
    \label{fig:strat}
\end{figure}

Both 2-D and 3-D simulations were conducted. The base grid resolution in the wave-normal ($x$) direction was 146.5 $\mu$m, and adding five AMR levels brought the minimum grid size to 4.58 $\mu$m. The simulations were conducted using roughly 1.3 million cells in 2-D and 1 billion cells in 3-D. Uniformly resolving the entire domain to the minimum grid sizes would have required $\sim$268 million cells in 2-D and $\sim$275 billion cells in 3-D, making them intractable without leveraging the AMR capabilities. A reference length scale for this problem is the induction length for the one-dimensional ZND detonation solution, which was computed as 654.6 $\mu$m for the mean reactant mixture conditions \cite{sdtoolbox}. Thus, the minimum grid size equates to roughly 143 cells per ZND induction length, which allowed the cellular instabilities at the wave fronts to be highly resolved. Refinement was added using pressure and density gradient criteria, which concentrated refinement on the leading and transverse shock waves, as well as the rapid heat release behind the waves. Representative snapshots along the channel mid-plane of one of the 3-D cases are shown in Fig.\ \ref{fig:strat}. Peaks in pressure, temperature, and heat release rate are observed following transverse wave collisions in detonable mixture regions (see $x\sim$65 mm, $y\sim$15 mm). However, when the wave traverses broad inert regions, the heat release rapidly decays, and the leading shock begins to decouple from the reaction front (see $x\sim$66 mm, $y>$23 mm). The localized strengthening and decay of the wave in the presence and absence of reactants, respectively, leads to greater variations in wave speed and broader reaction zones than in cases with uniform mixtures. The AMR methodology allows the local instabilities contributing to these differences in macroscopic wave dynamics to be efficiently captured. Further details on these studies can be found in Ref.\ \cite{ullman2024_strat}.

\clearpage
\bibliography{refs}

\clearpage
\appendix
\section{Instabilities in HLLC}\label{appendix:hllclm}
\begin{figure}[!htb]
\centering
      \includegraphics[width=.45\textwidth]{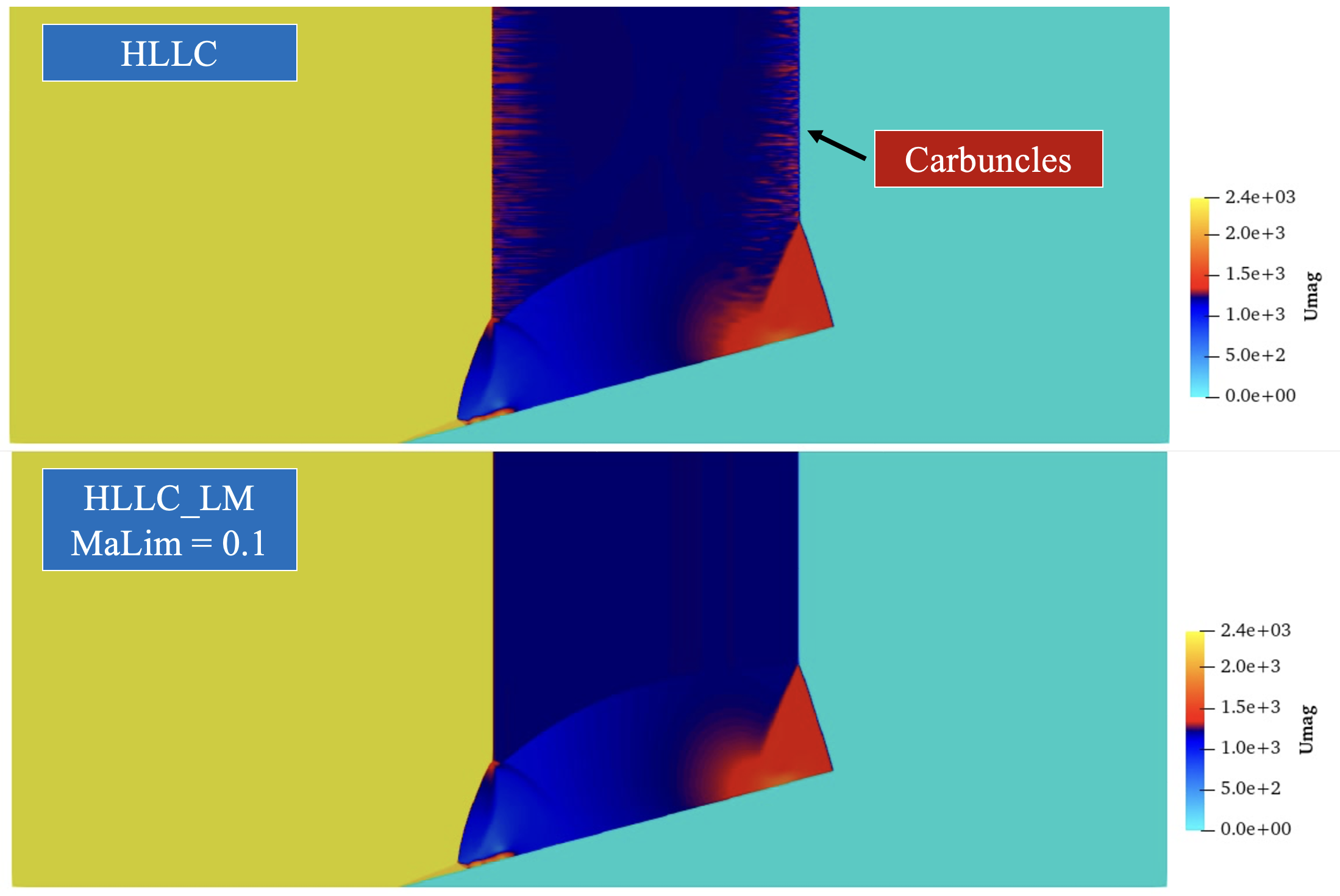}    
\caption{Inviscid Case: HLLC and HLLC-LM schemes.}
\label{fig.invs_scheme}
\end{figure}

The broad application of the contact-shear-preserving approximate Riemann solvers, such as the comprehensive wave schemes in the HLL family, is limited by the occurrence of various types of numerical shock instability. The prevailing perception is that this instability is fundamentally a multidimensional issue. This problem arises because these solvers that preserve contact and shear waves fail to provide sufficient cross-flow dissipation, which is necessary to suppress numerical fluctuations in physical variables near the shock wave. To highlight this problem, we simulate a 2D compression ramp at 15$^{\circ}$ with air inflow from the left at Mach 7 with pressure at 1 atm and temperature at 300 K. Figure~\ref{fig.invs_scheme} compares the results in the inviscid case in the same instance for the HLLC and HLLC-LM schemes. The presence of carbuncles can be seen at the shock interface in the top image as red streaks projecting along the flow direction. These instabilities grow as the simulation progresses and are present in cases such as standing shock. HLLC-LM shows a stable result due to the presence of an activation function as described in Section~\ref{sec:hllc-lm}.

\begin{figure}[!htb]
\centering
      \includegraphics[width=.45\textwidth]{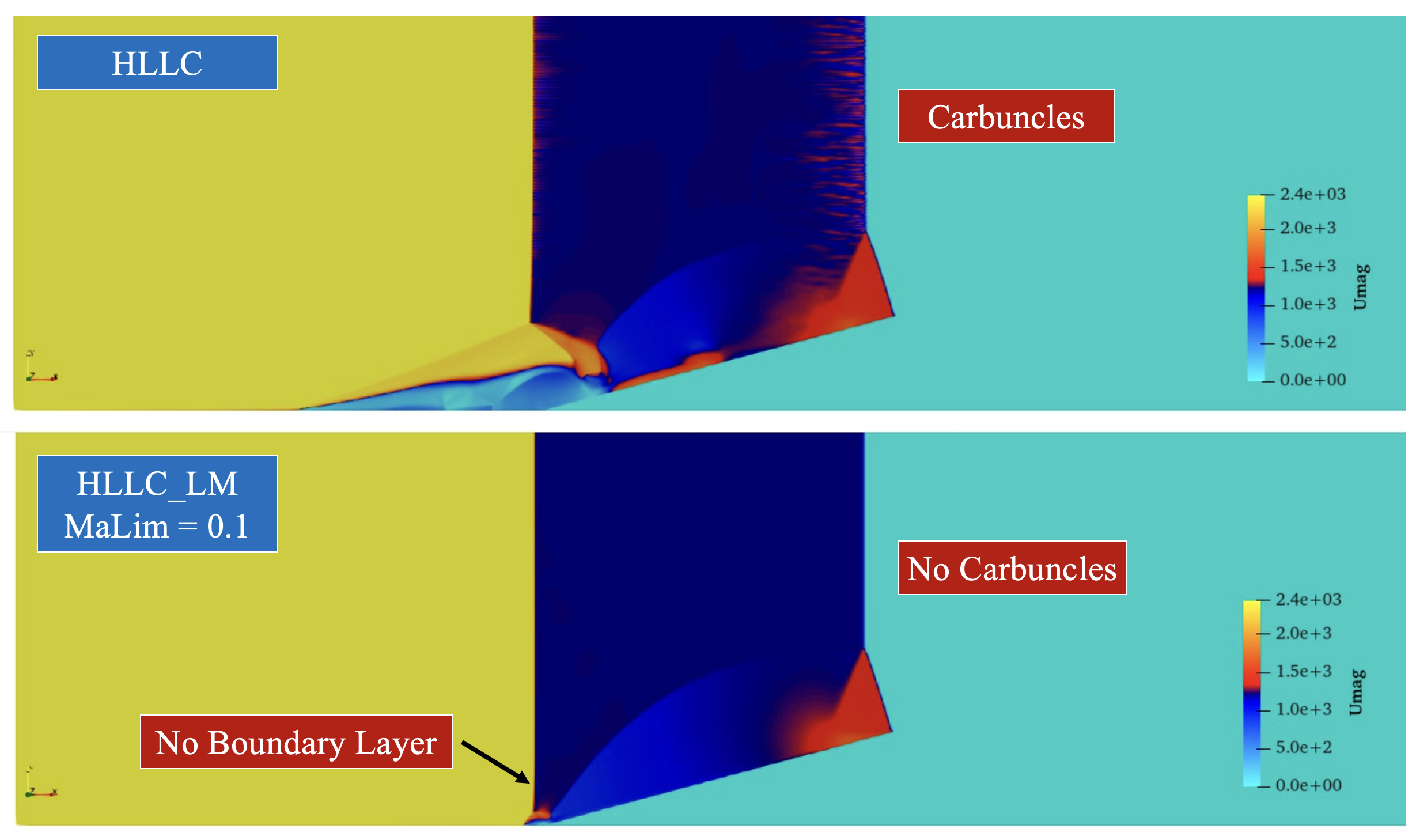}    
\caption{Viscous Case: HLLC and HLLC-LM schemes.}
\label{fig.visc_scheme}
\end{figure}

In viscous scenarios, the impact of the HLLC-LM scheme is different. As shown in Fig.~\ref{fig.visc_scheme}, the $M a_{\text {limit }} = 0.1$ tend to negatively affect the boundary layer (BL) while suppressing the carbuncles on the shock front. Clearly, BL is absent in the HLLC-LM case compared to the HLLC results. We perform experiments in various scenarios, including the forward-facing step, the backward-facing step, the compression ramp, and the 3D flat plate, adjusting the $M a_{\text{limit}}$ while considering the effects of viscosity, to investigate the role of $M a_{\text{limit}}$ as a ``limiting" parameter of the activation function $\phi$. We concluded that a value of $M a_{\text {limit }} \in [0.01 \sim 0.02]$ was able to limit carbuncles without affecting BL development. The HLLC-LM scheme can also be used directly with a limiter based on the approximate maximum speed of the carbuncle instability observed in the simulation. 

\end{document}